\documentclass[11pt]{article}
\usepackage{graphicx}
\usepackage{epsfig}

\newcommand{\BABARPubYear}    {00}

\newcommand{\BABARProcNumber} {14}
\newcommand{\SLACPubNumber} {8698}

\usepackage{relsize}
\def\babar{\mbox{\slshape B\kern-0.1em{\smaller A}\kern-0.1em
    B\kern-0.1em{\smaller A\kern-0.2em R}}}

\def\epem       {\ensuremath{e^+e^-}}

\def\mumu       {\ensuremath{\mu^+\mu^-}}

\def\s  {\ensuremath{s}}

\def\ccbar {\ensuremath{c\overline c}}
\def\b  {\ensuremath{b}}

\def\pipi  {\ensuremath{\pi^+\pi^-}}

\def\Kbar  {\kern 0.2em\overline{\kern -0.2em K}{}}

\def\KS    {\ensuremath{K^0_{\scriptscriptstyle S}}} 
\def\KL    {\ensuremath{K^0_{\scriptscriptstyle L}}}

\def\Kzb   {\ensuremath{\Kbar^0}}
\def\KzKzb {\ensuremath{K^0 \kern -0.16em \Kzb}}

\def\Dbar  {\kern 0.2em\overline{\kern -0.2em D}{}}

\def\Dzb   {\ensuremath{\Dbar^0}}
\def\DzDzb {\ensuremath{D^0 {\kern -0.16em \Dzb}}}

\def\Dstarb   {\ensuremath{\Dbar^*}}

\def\Dstarzb  {\ensuremath{\Dbar^{*0}}}

\def\Bz    {\ensuremath{B^0}}
\def\B     {\ensuremath{B}}
\def\Bbar  {\kern 0.18em\overline{\kern -0.18em B}{}}

\def\Bzb   {\ensuremath{\Bbar^0}}

\def\BzBzb {\ensuremath{B^0 {\kern -0.16em \Bzb}}}

\def\jpsi  {\ensuremath{{J\mskip -3mu/\mskip -2mu\psi\mskip 2mu}}} 
\def\psitwos {\ensuremath{\psi{(2S)}}}
\mathchardef\Upsilon="7107
\def\Y#1S{\ensuremath{\Upsilon{(#1S)}}}

\def\FourS {\Y4S}

\mathchardef\Deltares="7101
\mathchardef\Xi="7104
\mathchardef\Lambda="7103
\mathchardef\Sigma="7106
\mathchardef\Omega="710A
\def\Deltabar   {\kern 0.25em\overline{\kern -0.25em \Deltares}{}}
\def\Lbar {\kern 0.2em\overline{\kern -0.2em\Lambda\kern 0.05em}\kern-0.05em{}}
\def\Sigbar{\kern 0.2em\overline{\kern -0.2em \Sigma}{}}
\def\Xibar{\kern 0.2em\overline{\kern -0.2em \Xi}{}}
\def\Obar{\kern 0.2em\overline{\kern -0.2em \Omega}{}}
\def\Nbar{\kern 0.2em\overline{\kern -0.2em N}{}}
\def\Xbar{\kern 0.2em\overline{\kern -0.2em X}{}}

\def\ev   {\ensuremath{\rm \,e\kern -0.08em V}}
\def\kev  {\ensuremath{\rm \,ke\kern -0.08em V}} 
\def\mev  {\ensuremath{\rm \,Me\kern -0.08em V}} 
\def\gev  {\ensuremath{\rm \,Ge\kern -0.08em V}} 
\def\gevc {\ensuremath{{\rm \,Ge\kern -0.08em V\!/}c}} 
\def\tev  {\ensuremath{\rm \,Te\kern -0.08em V}}
\def\mevc {\ensuremath{{\rm \,Me\kern -0.08em V\!/}c}} 
\def\gevcc{\ensuremath{{\rm \,Ge\kern -0.08em V\!/}c^2}} 
\def\mevcc{\ensuremath{{\rm \,Me\kern -0.08em V\!/}c^2}}

\def\cm   {\ensuremath{\rm \,cm}}

\def\mm   {\ensuremath{\rm \,mm}}

\def\mum  {\ensuremath{\,\mu\rm m}} 

\def\invfb   {\ensuremath{\mbox{\,fb}^{-1}}}
\def\mus  {\ensuremath{\rm \,\mus}}

\def\ps   {\ensuremath{\rm \,ps}}

\def\mus        {\ensuremath{\,\mu{\rm s}}}    
\def\ps         {\ensuremath{{\rm \,ps}}}   
\def\gsim{{~\raise.15em\hbox{$>$}\kern-.85em
          \lower.35em\hbox{$\sim$}~}}
\def\lsim{{~\raise.15em\hbox{$<$}\kern-.85em
          \lower.35em\hbox{$\sim$}~}}

\def\CP                 {\ensuremath{C\!P}}

\def\to                 {\ensuremath{\rightarrow}}

\def\pep2{PEP-II}
\def\BF{$B$ Factory}
\def\abf {asymmetric \BF}

\newcommand{\dedx}{\ensuremath{\mathrm{d}\hspace{-0.1em}E/\mathrm{d}x}}

\def\sb{${\sin\! 2 \beta   }$}

\def\stwoa{\ensuremath{\sin\! 2 \alpha  }}
\def\stwob{\ensuremath{\sin\! 2 \beta   }}

\def\mistag{\ensuremath{w}}

\def\deltaz{\ensuremath{{\rm \Delta}z}}
\def\deltat{\ensuremath{{\rm \Delta}t}}
\def\deltamd{\ensuremath{{\rm \Delta}m_d}}

\newcommand{\eqref}[1]{Eq.~(\ref{eq:#1})}

\newcommand{\epjc}      [1]  {{Eur.\ Phys.\ Jour.\ C~{\bf #1}}}

\newcommand{\pl}        [1]  {{Phys.\ Lett.\ {\bf #1}}}      

\newcommand{\prl}       [1]  {{Phys.\ Rev.\ Lett.\ {\bf #1}}} 
\newcommand{\pr}        [1]  {{Phys.\ Rev.\ {\bf #1}}}

\def\jetset74   {\mbox{\tt Jetset \hspace{-0.5em}7.\hspace{-0.2em}4}}

\long\def\inst#1{\par\nobreak\kern 4pt\nobreak
    {\it #1}\par\vskip 10pt plus 3pt minus 3pt}

\begin{document}
{\pagestyle{empty}

\begin{flushright}
SLAC-PUB-\SLACPubNumber \\
\babar-PROC-\BABARPubYear/\BABARProcNumber \\
August, 2000 \\
\end{flushright}

\par\vskip 3cm

\begin{center}
\Large \bf 
 First {\boldmath \CP} Violation Results from \babar\ 
\end{center}
\bigskip

\begin{center}
\large 
David G. Hitlin\\
(for the \babar\ Collaboration)\\
California Institute of Technology\\ 
Pasadena, CA 91125 USA\\
E-mail: hitlin@hep.caltech.edu
\end{center}
\bigskip \bigskip

\begin{center}
\large \bf Abstract
\end{center}
We present a preliminary measurement of time-dependent 
\CP-violating asymmetries in  
$\Bz \to \jpsi \KS$ and $\Bz \to \psitwos \KS$  
decays recorded by the \babar\ detector at the \pep2\
asymmetric \BF\ at SLAC.  
The data sample consists of 9.0\invfb\  collected 
at the \FourS\ resonance and 0.8\invfb\ off-resonance.  
One of the neutral \B\ mesons, produced in pairs at the \FourS, is 
fully reconstructed.
The flavor of the other neutral \B\
meson is tagged at the time of its decay,  mainly with the charge of 
identified leptons and kaons.  
The time difference between the decays is determined 
by measuring the 
distance between the decay vertices.
Wrong-tag probabilities and the time resolution function
are measured 
with samples of fully-reconstructed semileptonic 
and hadronic neutral \B\ final states.
The value of the asymmetry amplitude, \stwob, is determined from
a maximum likelihood fit 
to the time distribution of 120 tagged 
$\Bz \to \jpsi \KS$ and $\Bz \to \psitwos \KS$  candidates:
\stwob=0.12 $\pm$ 0.37 {\rm (stat)} $\pm$ 0.09 {\rm (syst)}. }

\vfill
\begin{center}
Contributed to the Proceedings of the 30$^{th}$ International 
Conference on High Energy Physics, \\
7/27/2000---8/2/2000, Osaka, Japan
\end{center}

\vspace{1.0cm}
\begin{center}
{\em Stanford Linear Accelerator Center, Stanford University, 
Stanford, CA 94309} \\ \vspace{0.1cm}\hrule\vspace{0.1cm}
Work supported in part by Department of Energy contract DE-AC03-76SF00515.
\end{center}

\setlength\columnsep{0.20truein}
\twocolumn
\def\sloppy{\tolerance=100000\hfuzz=\maxdimen\vfuzz=\maxdimen}
\sloppy
\vbadness=12000
\hbadness=12000
\flushbottom
\def\figurebox#1#2#3{%
  	\def\arg{#3}%
  	\ifx\arg\empty
  	{\hfill\vbox{\hsize#2\hrule\hbox to #2{\vrule\hfill\vbox to #1{\hsize#2\vfill}\vrule}\hrule}\hfill}%
  	\else
   	{\hfill\epsfbox{#3}\hfill}%
  	\fi}

\section{Introduction}
\label{sec:Introduction}
The \babar\ detector at the \pep2\ \abf\ at SLAC has been taking data since the end of May, 1999.
As of the date of this conference a data sample of $\sim$9.0\invfb\  has been collected 
at the \FourS\ resonance, with an additional 0.8\invfb\ taken off-resonance. A variety of preliminary
$B$ physics results using this data sample have been submitted to this conference and reported in 
parallel sessions. This paper will discuss in some detail \babar's first measurements of \CP-violating 
asymmetries in $\Bz \to \jpsi \KS$ and $\Bz \to \psitwos \KS$ decays. 

\section{Motivation and Overview}
\hbox{The \CP-violating phase of the three-genera-} tion Cabbibo-Kobayashi-Maskawa (CKM)
quark mixing matrix can provide an elegant explanation of the well-established \CP-violating
effects seen in \KL\ decay\cite{EpsilonK}.  
However, studies of \CP\ violation 
in neutral kaon decays and the resulting experimental constraints on the parameters of the CKM
matrix\cite{MSConstraints} do not, in fact, yet provide a test of whether the CKM phase describes
\CP\ violation\cite{Primer}. 

The unitarity of the three generation CKM matrix can be expressed in geometric form as six
triangles of equal area in the complex plane. 
A nonzero area\cite{Jarlskog} 
directly implies the existence of a \CP-violating CKM phase. The most experimentally accessible
of the unitarity relations, involving the two smallest elements of the CKM matrix, $V_{ub}$ and
$V_{td}$, has come to be known as the ($B$)
Unitarity Triangle. 
Because the lengths of the sides of 
this  Unitarity Triangle are of the same order, the angles  
can be large, leading to potentially large \CP-violating 
asymmetries 
from phases between CKM matrix elements.

The \CP-violating asymmetry in  $\b \to \ccbar \s$ decays of the \Bz\ meson 
such as  $\Bz/\Bzb \to \jpsi \KS$ 
(or $\Bz/\Bzb \to \psitwos \KS$) 
is caused by the interference between mixed and unmixed decay amplitudes.
A state initially prepared as a \Bz\ (\Bzb) can decay directly to $\jpsi \KS$ 
or can oscillate into a \Bzb\ (\Bz) and then decay
to $\jpsi \KS$.  With little theoretical uncertainty, the phase difference 
between these amplitudes is equal to twice the angle 
$\beta = \arg \left[\, -V_{cd}V_{cb}^* / V_{td}V_{tb}^*\, \right]$ of the
Unitarity Triangle.  The \CP-violating asymmetry can thus provide a crucial test of the Standard Model. 
The interference between the 
two amplitudes, and hence the \CP\ asymmetry, is maximal 
when the mixing probability is at its highest, {\it i.e.}, when the lifetime $t$ is approximately 
2.2 \Bz\ proper lifetimes.

In \epem\ storage rings operating at  the \FourS\ resonance a \BzBzb\ pair
produced in \FourS\ decay 
evolves in a coherent $P$-wave until one of the \B\ mesons decays. 
If one of the \B\ mesons ($B_{tag}$) can be 
ascertained to decay to a state of known flavor at a certain time $t_{tag}$, 
the other \B\ is {\it at that time} known to be of the opposite flavor.
For this measurement, the other \B\ ($B_{CP}$) is fully reconstructed in a \CP\ 
eigenstate ($\jpsi \KS$ or $\psitwos \KS$).
By measuring the proper time interval $\deltat = t_{CP} - t_{tag}$ from the $B_{tag}$ decay time 
to the decay of the $B_{CP}$, it is possible to determine the time evolution 
of the initially pure \Bz\ or \Bzb\ state.  
The time-dependent rate of decay of the $B_{CP}$ final state is given by 
\begin{eqnarray}
 f_\pm(\, \deltat \, ; \,  \Gamma, \, \deltamd, \, {\cal {D}} \sin{ 2 \beta } )  =
{\frac{1}{4}}\, \Gamma \, {\rm e}^{ - \Gamma \left| \deltat \right| }  \nonumber \\
\!\!\!\!\!\!\!\!\!\!\!\!\!\!\!\!\!\! \times  \left[ \, 1 \, \pm  \,{\cal {D}} \sin{ 2 \beta } \times \sin{ \deltamd  \deltat }   \right],\ \ 
\label{eq:TimeDep}
\end{eqnarray}
where the $+$ or $-$  sign 
indicates whether the 
$B_{tag}$ is tagged as a \Bz\ or a \Bzb, respectively.  The dilution factor ${\cal {D}}$ is given by 
$ {\cal {D} } = 1 - 2 \mistag$, where $\mistag$ is the mistag fraction, {\it i.e.}, the 
probability that the 
flavor of the tagging \B\ is identified incorrectly. 
A term proportional to $\cos {  \deltamd \, \deltat }$ 
would arise from the interference between two decay mechanisms with different
weak phases.  In the Standard Model,
the dominant diagrams (tree and penguin) for the decay  modes we consider have no relative weak phase, so no such term is expected.

To account for the finite resolution of the detector,
the time-dependent distributions $f_\pm$ for \Bz\ and \Bzb\ tagged events 
(Eq.~\ref{eq:TimeDep}) must be convoluted with 
a time resolution function ${\cal {R}}( \deltat ; \hat {a} )$:

\begin{eqnarray}
{\!\!\!\!\!\!\!\!\!\!\!\! \cal F}_\pm(\, \deltat \, ; \, \Gamma, \, \deltamd, \, {\cal {D}} 
\, \sin{ 2 \beta }, \hat {a} )  =  \;\;\;\;\;\;\;\;\;\;\;\; \nonumber \\
f_\pm(  \deltat  ;  \Gamma,  \deltamd,  {\cal {D}}\, \sin{ 2 \beta }  ) \otimes 
{\cal {R}}(  \deltat  ;  \hat {a}  ) , \label{eq:Convol}
\end{eqnarray}
where $\hat {a}$ represents the set of parameters that describe the resolution function. 

In practice, events are separated into different tagging categories, 
each of which has a different mean dilution ${\cal {D}}_i$, determined 
individually for each category.
\par
It is possible to construct a \CP-violating observable 
\begin{equation}
  {\cal A}_{CP}(\deltat) = \frac{ {\cal F}_+(\deltat) \, - \, {\cal F}_-(\deltat) }
{ {\cal F}_+(\deltat) \, + \, {\cal F}_-(\deltat) } \ \ , 
\label{eq:asymmetry}
\end{equation}
which is proportional to \stwob:
\begin{equation}
 {\cal A}_{CP}(\deltat) \sim {\cal D} \sin{ 2 \beta } \times \sin{ \deltamd \, \deltat } \ \ .
\label{eq:asymmetry2}
\end{equation}

Since no time-integrated \CP\ asymmetry effect is expected, an
analysis of the time-dependent asymmetry is necessary.  
At an asymmetric-energy \BF, the proper decay-time difference $\deltat$ is, 
to an excellent approximation, proportional to 
the distance \deltaz\ between  the two \Bz-decay vertices 
along the axis of the boost, 
$\deltat \approx \deltaz / {\rm c} \left< \beta \gamma \right>  $.  
At \pep2 the average boost of \B\ mesons, 
$\left< \beta \gamma \right>$, is $0.56$. 
The distance $\deltaz$ is  250 $\mum$ per \Bz\ lifetime, while the typical 
$\deltaz$ resolution for the \babar\ detector is about 
$110 \mum$. 

Since the amplitude of the time-dependent \CP -violating 
asymmetry in Eq.~\ref{eq:asymmetry2} 
involves the product of ${\cal {D}}$ and \stwob, one needs to determine 
the dilution factors ${\cal {D}}_i $ (or equivalently the mistag fractions $\mistag_i$) 
in order to extract the value of \stwob.
The mistag fractions can be extracted from
the data
by studying the time-dependent rate of \BzBzb\ oscillations in events
in which one of the neutral \B\ mesons is fully reconstructed in a 
self-tagging mode
and the other \B\ (the $B_{tag}$) is flavor-tagged using 
the standard \CP\ analysis flavor-tagging algorithm.
In the limit of perfect determination 
of the flavor of the fully-reconstructed neutral \B, the dilution in the 
mixed and unmixed amplitudes arises solely from the $B_{tag}$ side, 
allowing the values of the mistag fractions $\mistag_i$ to be determined. 

The value of \stwob\ is extracted by maximizing the likelihood function
The value of \stwob\ is extracted by maximizing the likelihood function
\begin{eqnarray}
& \ln {\cal {L} }_{C\!P}= \;\;\;\;\;\;\;\;\;\;\;\;\;\;\;\;\;\;\;\;\;\;\;\;\;\;\;\;\;\;\;\;\;
\;\;\;\;\;\;\;\;\;\;\;\;\;\;\;\;\;\;\;\;\;\;\;\;\;\;\;\;     \\
 & \sum_{ i}
\bigg[  \sum_{ \Bz{\rm tag} }  \ln{ {\cal F}_+(  \deltat  ;   \Gamma,  
\deltamd, \, \hat {a}, 
\, {\cal {D}}_i \sin{ 2 \beta }  ) } \;\;\;\;\;\; \nonumber\\
 & +  \sum_{ \Bzb  {\rm tag} } { \ln{ {\cal F}_-(  \deltat  ;   \Gamma,  
\deltamd, \,  \hat {a}, \, {\cal {D}}_i \sin{ 2 \beta }  \, ) } } \bigg], \nonumber 
\end{eqnarray}
where the outer summation is over tagging categories $i$.

\subsection{Overview of the analysis}
\label{sec:Overview}
The measurement of the \CP-violating asymmetry has five main components~:
\begin{itemize}
\item
Selection of the signal $\Bz/\Bzb \to \jpsi \KS$ 
and $\Bz/\Bzb \to \psitwos \KS$ events, as described in detail
in Ref.~5.
\item
Measurement of the distance \deltaz\ between the two \Bz\ decay 
vertices along the \FourS\ boost axis, as described in detail in Refs.~6 and~7.
\item
Determination of the flavor of the $B_{tag}$, as described 
in detail in Ref.~6.
\item 
Measurement of the dilution factors ${\cal D}_i$ from the data  
for the different tagging categories, as described in detail
in Ref.~6.
\item
Extraction of the amplitude of the \CP\ asymmetry and the value of \stwob\
with an unbinned maximum likelihood fit.  
\end{itemize}
\par
Whenever possible, we determine time and mass resolutions, efficiencies and 
mistag fractions from the data.  

\section{Sample selection} 
\label{sec:Sample}
\par
For this analysis we use a sample of $9.8 \invfb$ of data recorded
by the \babar\ detector\cite{BabarPub0018} 
between January 2000 and the beginning of July 2000, 
of which $0.8 \invfb$ was recorded 40\mev\ below the \FourS\ resonance (off-resonance data). 
\par 
A brief description of the \babar\ detector and the definition of many 
general analysis procedures can be found in Ref.~8. 
Charged particles
are detected and their momenta measured by a combination of 
a central drift chamber (DCH) filled with a helium-based gas and  a
five-layer, doubled-sided silicon vertex tracker (SVT), 
in a 1.5~T solenoidal field produced by a superconducting magnet.  The 
charged particle momentum resolution is approximately 
$\left( \delta p_T / p_T \right)^2 =  (0.0015\, p_T)^2 + (0.005)^2$, where 
$p_T$ is measured in \gevc.  The SVT, 
with typical 10\mum\ single-hit resolution, 
provides vertex information in both  
the transverse plane and in the $z$ direction.  
Vertex resolution is typically 50\mum\ in $z$  for 
a fully reconstructed \B\ meson, depending on the decay mode, and of order 
100 to 150\mum\ for a generic \B\ decay.
Leptons and hadrons are identified with measurements
from all the \babar\ components, including 
the energy loss $dE/dx$ from a truncated mean of up to
40 samples in the DCH and at least 8 samples in the SVT.
Electrons and photons are identified in the barrel and the forward regions 
by the CsI electromagnetic calorimeter (EMC). Muons are identified in the
instrumented flux return (IFR).   In the central polar region the Cherenkov ring
imaging detector (DIRC) provides $K$-$\pi$ separation 
with a significance of at least three standard deviations over the full 
momentum range for $B$ decay products above 250\mevc.

\subsection{Particle identification}

An electron candidate must be matched to an electromagnetic cluster 
of at least three crystals in the CsI calorimeter. 
The ratio of 
the cluster energy to the track momentum, 
$E/p$,  must be between 0.88 and 1.3.  
The lateral moment of the cluster must be between 0.1 and 0.6, 
and  the Zernike moment of order (4,2)\footnote{The lateral and Zernike moments are cluster 
shape variables introduced in Ref.~8.} 
must be  smaller than 0.1. 
In addition the 
electron candidate track in the drift chamber must have a $dE/dx$ measurement 
consistent with that of an electron and, if measured, 
the Cherenkov angle in the 
DIRC must be consistent with that of an ultra-relativistic particle.

Muon identification relies principally on 
the measured number of interaction 
lengths, $N_\lambda$, penetrated by the candidate in the IFR iron, 
which must have a minimum value of $2.2$ and,
at higher momenta, must be 
larger than
$N_\lambda^{exp}-1$, where $N_\lambda^{exp}$ is the expected number of interaction 
lengths for a muon.  
The number of IFR layers with a ``hit'' must be larger than two. 
To reject hadronic showers, 
we impose criteria on the number of IFR strips with a hit as a function of the 
penetration length, and on the distance between the strips with hits and the extrapolated track.
In the forward region, which suffers from accelerator-related background, 
extra hit-continuity criteria are applied.  
In addition,  if the muon candidate is in the angular region 
covered by the EMC,
the energy deposited by the candidate in the calorimeter must be 
larger than 50\mev\ and
smaller than 400\mev. (The expected energy deposited by a minimum 
ionizing particle is about 180\mev.)

Particles are identified as kaons if the ratio 
of the combined kaon likelihood 
to the combined pion likelihood is greater than 15.   
The combined likelihoods are the product of the individual 
likelihoods 
in the SVT, DCH and DIRC subsystems.
In the SVT and DCH tracking detectors, the likelihoods 
are based on the measured  $dE/dx$ truncated mean 
compared to the expected 
mean for the $K$ and $\pi$ hypotheses, with an assumed 
Gaussian distribution.
The $dE/dx$ resolution is estimated
on a track-by-track basis, based on the direction and momentum of the track and the 
number of energy deposition samples.
For the DIRC, the likelihood is computed by combining the likelihood  of the measured Cherenkov
angle compared to  the expected Cherenkov angle for a given hypothesis, 
with the Poisson probability of the number of observed Cherenkov photons, 
given the number of expected 
photons for the same  hypothesis.
DIRC information is not required for particles with momentum less than 0.7\gevc,  
where the 
DCH $dE/dx$  alone provides good $K / \pi$ discrimination.

\subsection{Data samples}
We define three event classes{\footnote{ Throughout this paper, conjugates  
of flavor-eigenstate modes are implied.}}:
\begin{itemize}
\item
the \CP\ sample, containing
\Bz\ candidates reconstructed in the \CP\ eigenstates
$\jpsi \KS$ or $\psitwos \KS$.  
The charmonium mesons
\jpsi\ and \psitwos\ are reconstructed through their decays to \epem\ 
and \mumu. The \psitwos\ is also reconstructed through its decay 
to $\jpsi\pipi$. The \KS\ is reconstructed through its decay to \pipi\
and $\pi^0 \pi^0$.
The selection criteria for the \CP\ sample are described in the next section.

\item
the fully reconstructed \Bz\ samples, containing  
\Bz\ candidates in either semileptonic or hadronic flavor eigenstates.
The sample of semileptonic decays contains candidates in the 
 $\Bz \to D^{* -} \ell^+ \nu_\ell$  
mode ($\ell^+=e^+$ or $\mu^+$);
the sample of hadronic neutral decays contains
\Bz\ candidates in the  $D^{(*)-} \pi^+$, 
 $D^{(*)-} \rho^+$ and  $D^{(*)-} a_1^+$ modes;
the sample of hadronic charged decays contains
$B^+$ candidates in the  $\Dzb \pi^+$, 
and  $\Dstarzb \pi^+$ (with $\Dstarzb \to \pi^0 \Dzb$) modes.
The selection criteria for these samples are described 
in Refs.~6 and 7.
We reconstruct $\approx 7500$  $\Bz \to D^{* -} \ell^+ \nu_\ell$  candidates, 
$\approx 2500$ candidates in hadronic \Bz\ final states, and 
$\approx 2300$ candidates in hadronic $B^+$ final states.
\item
the charmonium control samples,  containing fully reconstructed
neutral or charged \B\ candidates in two-body decay 
modes with a \jpsi\ in the final state, such as $B^+ \to \jpsi K^+$ 
or $\Bz \to \jpsi (K^{* 0} \to K^+ \pi^-)$.
The selection criteria for these samples are described in Ref.~5.
We reconstruct 570 $B^+ \to \jpsi K^+$ candidates and 237 $\Bz \to \jpsi (K^{* 0} \to K^+ \pi^-)$ candidates. 
\end{itemize}

Signal event yields and purities for the individual samples are summarized in
Table~\ref{tab:CharmoniumYield}.

\subsection{The \CP\ sample}

We select
events with a minimum of four reconstructed charged tracks 
in the region defined by $0.41 < \theta_{lab} < 2.41$.
Events are required to have a reconstructed vertex 
within 0.5\cm\ of the average position of the 
interaction point in the plane transverse to the beamline, 
and a total energy greater than 5\gev\ in the fiducial regions for 
charged tracks and neutral clusters.
To reduce continuum background, 
we require the second-order normalized Fox-Wolfram moment\cite{fox}  
($R_2=H_2/H_0$) of the event to be less than 0.5. 
\par
The selection criteria for the $\jpsi \KS$ and $\psitwos \KS$ events are 
optimized by maximizing the ratio ${\cal S}/\sqrt{ {\cal S}+{\cal B}}$,
where ${\cal S}$ (the number of signal events that pass the selection) 
is determined from signal Monte Carlo events, and ${\cal B}$ (the number of
background events that pass the selection) is estimated from
a luminosity-weighted average of continuum data events and nonsignal $\B \Bbar$ Monte Carlo events. 
\par 
For the $\jpsi$ or $\psitwos \to \epem$ candidates,
at least one of the decay products is required to be positively identified as an electron or,
if outside the acceptance of the calorimeter, to be consistent with an electron 
according to  the drift chamber \dedx\ information.  
If both tracks are within the calorimeter acceptance and have a 
value of $E/p$ larger than 0.5, an 
algorithm for the recovery of Bremsstrahlung photons\cite{BabarPub0005} is used.
\par
For the $\jpsi$ or $\psitwos \to \mu^+ \mu^-$ candidates, 
at least one of the decay products is required to be 
positively identified as a muon
and the other, if within the acceptance of the calorimeter, is required to be 
consistent with a  minimum ionizing particle. 
\par
We select $\jpsi$ candidates with an invariant mass 
greater than 2.95\gevcc\ and 3.06\gevcc\ 
for the $\epem$ and $\mu^+ \mu^-$ modes, respectively, 
and  smaller than 3.14\gevcc\ in both cases.  
The $\psitwos$ candidates in leptonic modes must have a mass 
within 50\mevcc\ of the  $\psitwos$ mass.
The lower bound is relaxed to 250\mevcc\ for the $\epem$ mode.

\par 
For the $\psitwos \to \jpsi \pi^+ \pi^-$ mode, mass-constrained \jpsi\ candidates 
are combined with pairs of oppositely charged 
tracks considered as pions, and  $\psitwos$ candidates with mass between 3.0\gevcc\ 
and 4.1\gevcc\ are retained.  
The mass difference between the \psitwos\ candidate
and the \jpsi\ candidate is required to be within 15\mevcc\ of the known
mass difference. 

\par
\KS\ candidates reconstructed in the $\pi^+ \pi^-$ mode are required to have
an invariant mass, computed at the vertex of the two tracks, 
between 486\mevcc\ and 510\mevcc\ for the $\jpsi \KS$ selection, 
and  between 491\mevcc\ and 505\mevcc\ for the $\psitwos \KS$ selection. 

For the  $\jpsi \KS$ mode, we also consider the decay of the \KS\ 
into $\pi^0 \pi^0$.   
Pairs of $\pi^0$ candidates, with total energy above 800\mev\ 
and invariant mass, measured at the primary vertex, between 300 and 
700\mevcc, are considered as \KS\ candidates.  For each candidate, 
we determine the most probable \KS\ decay point along the path defined 
by the \KS\ momentum vector and the primary vertex of the event.
The decay-point probability is the product of the $\chi^2$ probabilities 
for each photon pair constrained to the $\pi^0$ mass.
We require the distance from the decay point to the primary vertex 
to be between $-10$~cm and $+40$~cm and 
the \KS\ mass measured at this point to be between 470 and 536\mevcc.

\begin{figure}
\begin{center}
\epsfxsize2.7in
\figurebox{}{}{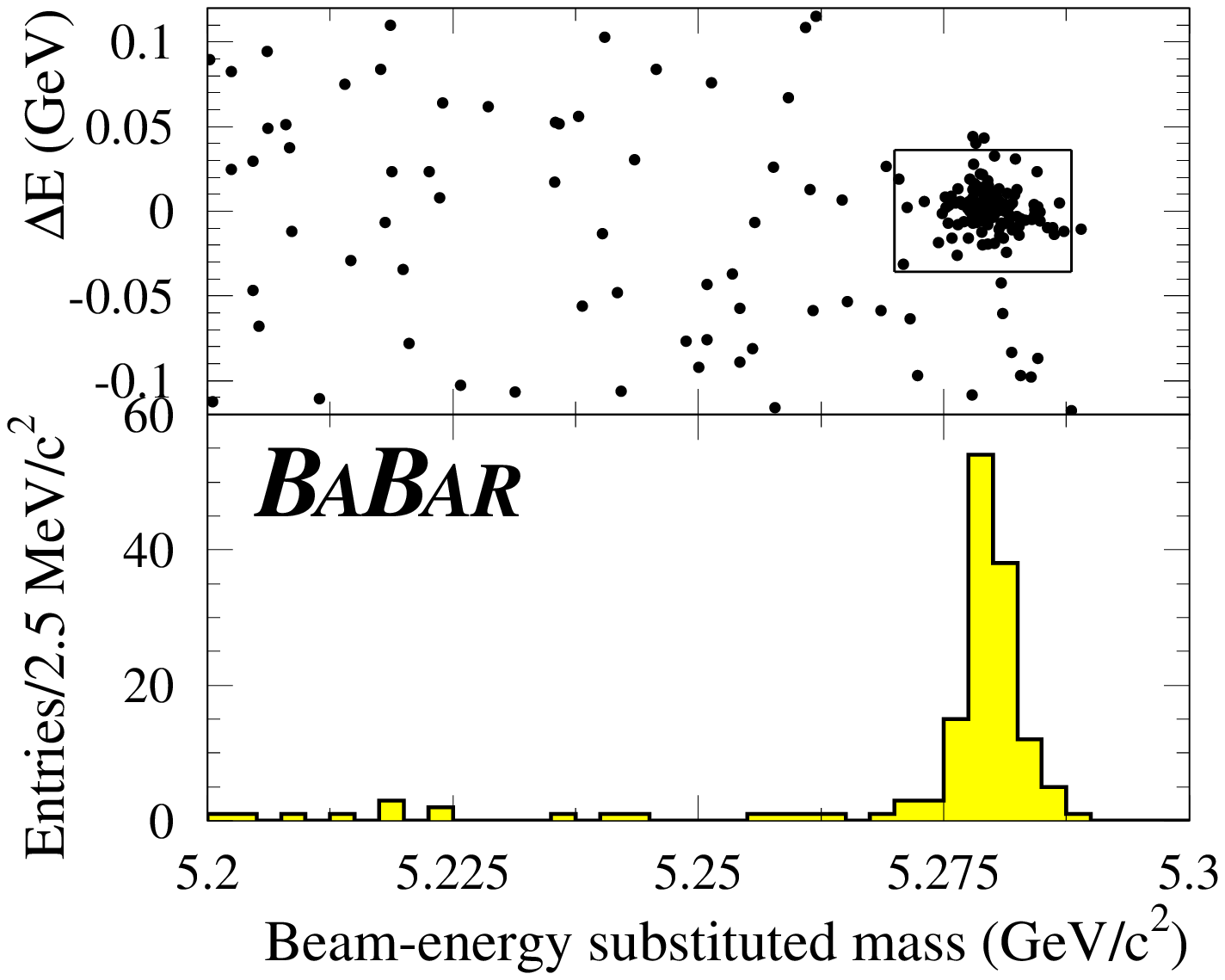}
\caption{$\jpsi \KS$ ($\KS \to \pi^+ \pi^-$) signal.}
\label{fig:jkspm}

\figurebox{}{}{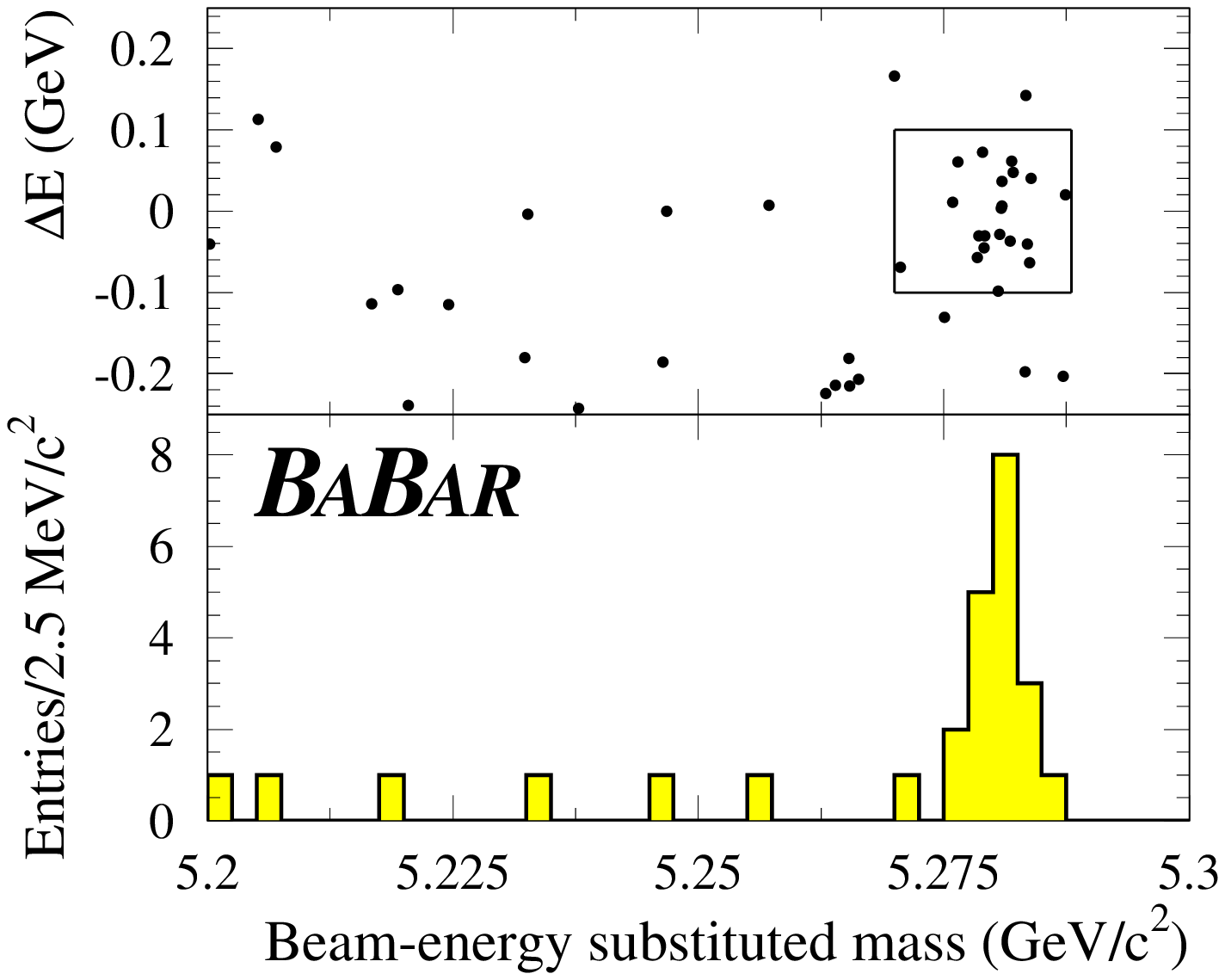}
\caption{$\jpsi \KS$ ($\KS \to \pi^0 \pi^0$) signal.}
\label{fig:jks00}

\figurebox{}{}{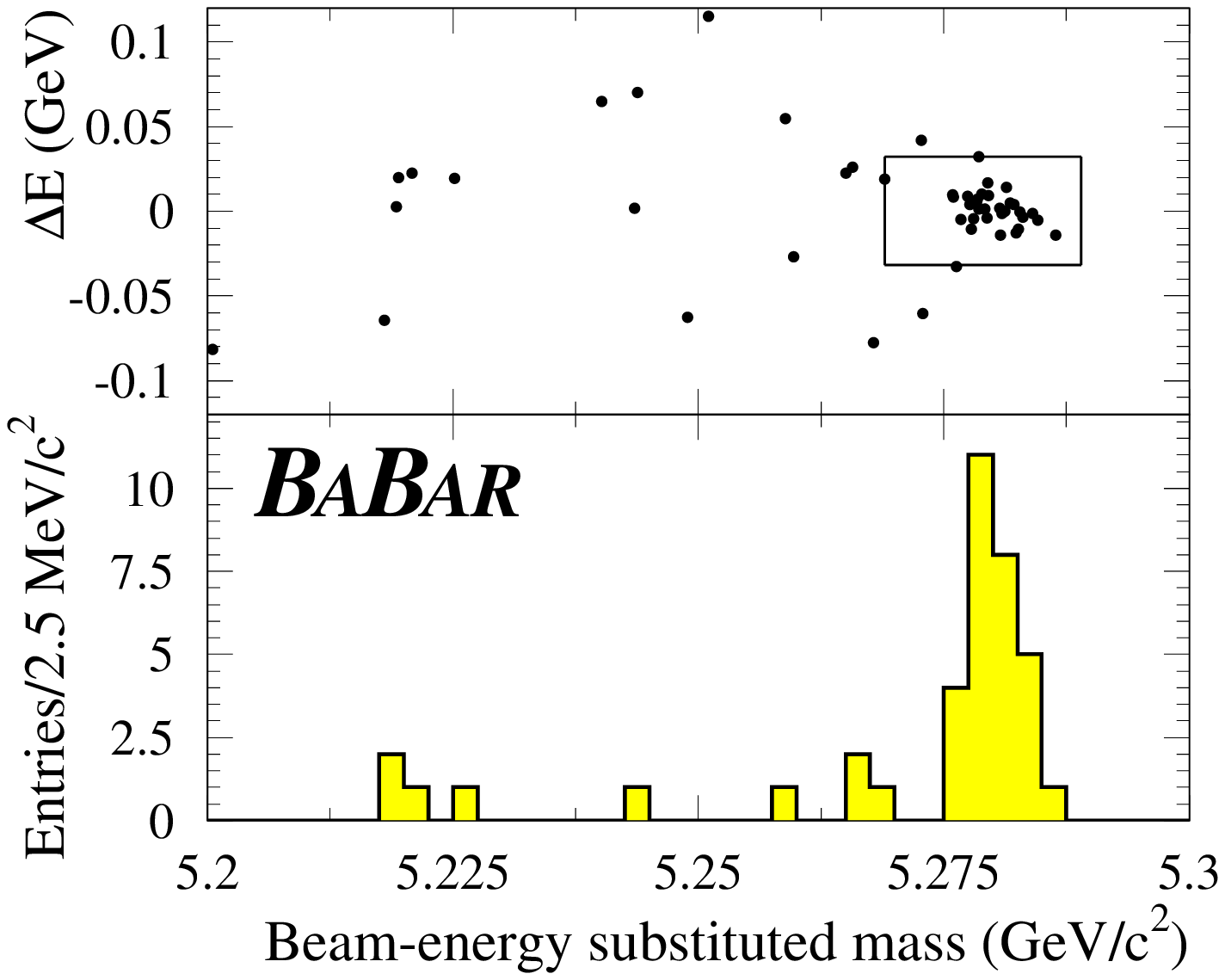}
\caption{$\psitwos \KS$ ($\KS \to \pi^+ \pi^-$) signal.}
\label{fig:psi2sks}
\end{center}
\end{figure}

\begin{figure}[!t]
\begin{center}
\epsfxsize2.7in

\figurebox{}{}{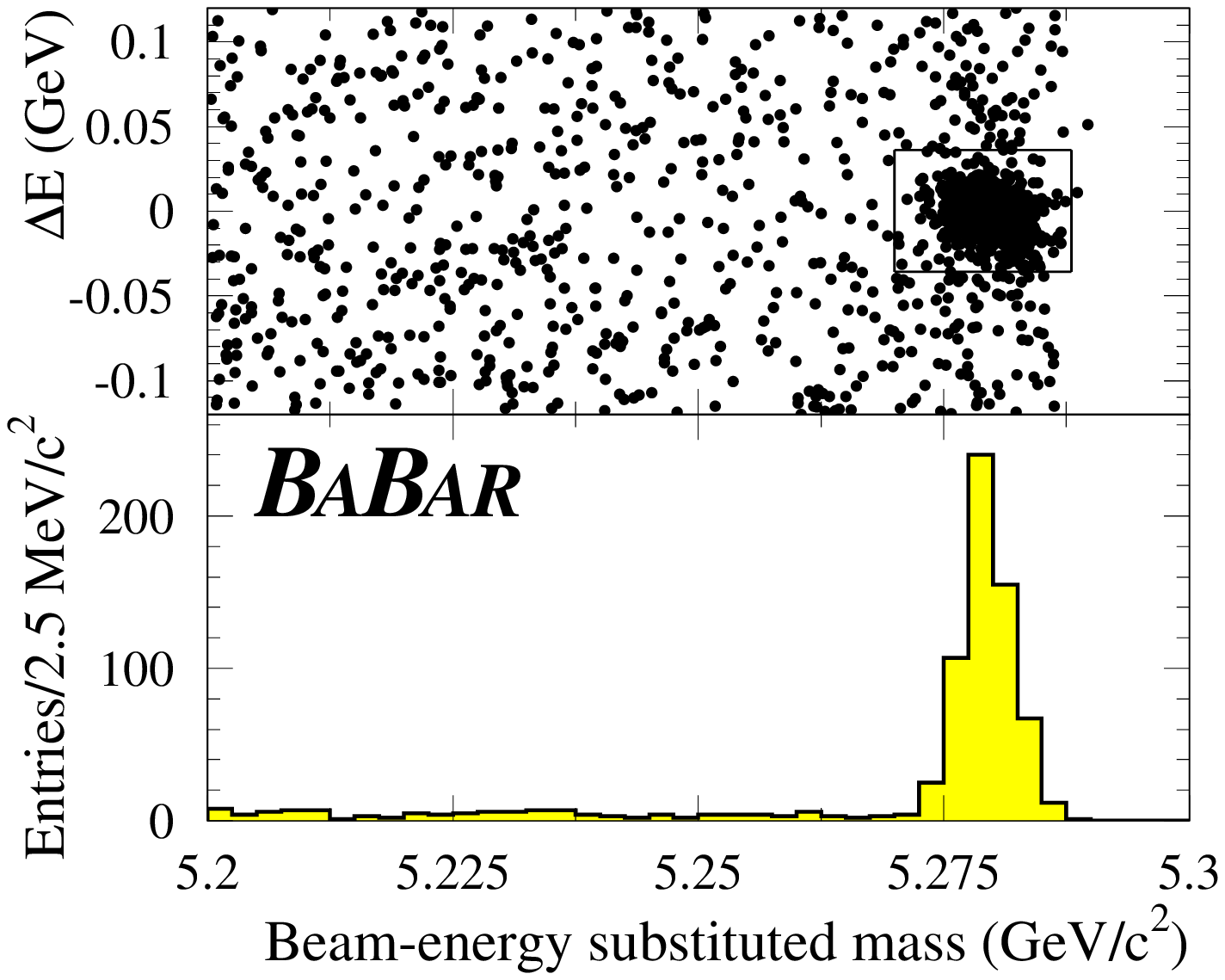}
\caption{$\jpsi K^+$ signal.}
\label{fig:jkskpm}

\figurebox{}{}{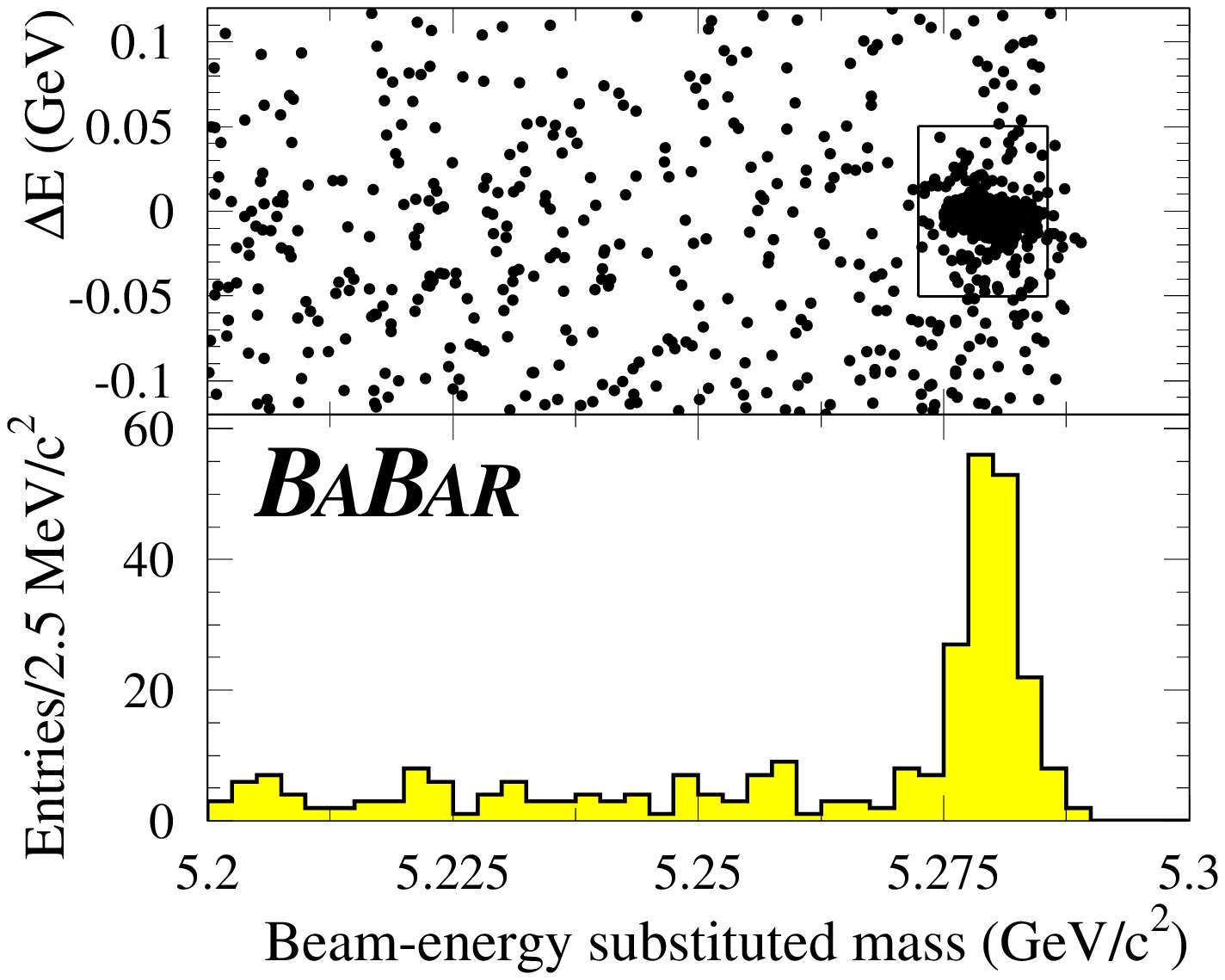}
\caption{$\jpsi K^{*0}$ ($K^{*0} \to K^+ \pi^-$) signal.}
\label{fig:jkstpm}

\end{center}
\end{figure}

$B_{CP}$ candidates are formed by combining mass-constrained \jpsi\ or
\psitwos\ candidates with mass-constrained \KS\ candidates.
The cosine of the angle between  the \KS\ 
three-momentum vector and  the vector 
that links the \jpsi\ and \KS\ vertices must be positive. 
The cosine of the helicity angle of the \jpsi\ in the \B\ 
rest frame
must be less than 0.8 for the $\epem$ mode and 0.9 for the 
$\mu^+ \mu^-$ mode. 
  
\par  
For the $\psitwos \KS$ candidates, the helicity angle
of the $\psitwos$ must be smaller than 0.9 for both leptonic modes.
The \KS\ flight length with respect to the $\psitwos$ vertex 
is required to be greater than 1\mm.
In the $\psitwos \to \jpsi \pi^+ \pi^-$ mode, the absolute value of the 
cosine of the 
angle between the $B_{CP}$ candidate three-momentum vector and the 
thrust vector of the rest of the event, in the center-of-mass 
frame, must be less than 0.9.

\par 
$B_{CP}$ candidates are identified with a pair of nearly uncorrelated 
kinematic variables: the 
difference ${\rm \Delta}E$ between the energy of the $B_{CP}$ candidate and the 
beam energy in the center-of-mass frame, 
and the beam-energy substituted mass\cite{BabarPub0018} $m_{\rm SE}$.
The signal region is defined by 
$ 5.270 \gevcc < m_{\rm SE} <  5.290 \gevcc$ and
an approximately three-standard-deviation cut on ${\rm \Delta}E$
(typically $\left| {\rm \Delta} E \right| < 35\mev$).

\par
Distributions of ${\rm \Delta} E$ and $m_{\rm SE}$ are shown 
in Fig.~\ref{fig:jkspm}, \ref{fig:jks00} and~\ref{fig:psi2sks} for the \CP\ samples and in 
Fig.~\ref{fig:jkskpm} and~\ref{fig:jkstpm} for the charmonium control samples. 
Signal event yields and purities, determined 
from a fit to the $m_{\rm SE}$ distributions after selection 
on ${\rm \Delta} E$, are summarized in
Table~\ref{tab:CharmoniumYield}. 

\par
The \CP\ sample used  in this 
analysis is composed of 168 candidates:  
121 in the $\jpsi \KS$ ($\KS \to \pi^+ \pi^-$) channel, 
19 in the  $\jpsi \KS$ ($\KS \to \pi^0 \pi^0$) channel and 
28 in the $\psitwos \KS$ ($\KS \to \pi^+ \pi^-$) channel. 

\section{Time resolution function}
\par
The resolution of the \deltat\ measurement 
is dominated by the $z$ resolution of the tagging vertex.
The tagging vertex is determined as follows. 
The three-momentum of the tagging $B$ and its associated 
error matrix are derived from the fully reconstructed 
 $B_{CP}$ candidate three momentum, 
decay vertex and
error matrix, and from the knowledge of the average position
of the interaction point and the \FourS\ four-momentum.
This derived $B_{tag}$ three-momentum is fit to a common vertex with the 
remaining tracks in the event (excluding those from $B_{CP}$).
In order to reduce 
the bias due to long-lived particles, all reconstructed $V^0$ 
candidates are used as input to the fit in place
of their daughters.  Any track whose contribution to the $\chi^2$
is greater than 6 is removed from the fit. 
This procedure is iterated until there
are no tracks contributing more than 6 to the $\chi^2$ 
or until all tracks are removed.   
Events are rejected if the fit does not converge for either the $B_{CP}$ or 
$B_{tag}$ vertex.
We also reject events with large \deltaz\ 
( $\left| \deltaz \right| > 3\mm$) or a large error on \deltaz\
($\sigma_{\deltaz}>400\mum$).

\begin{table*}
\caption{
Event yields for the different samples used in this analysis, 
from the fit to  $m_{\rm SE}$ distributions after selection 
on ${\rm \Delta} E$.  The purity is quoted for 
$ m_{\rm SE}>5.270 \mevcc$ (except for  $D^{*-}\ell^+\nu$). 
} 
\vspace{0.3cm}
\begin{center}
\begin{tabular}{|l|l|c|c|} \hline
Sample &  Final state  & Yield & Purity (\%) \\ \hline \hline
\CP & $\jpsi \KS$ ($\KS \to \pi^+\pi^-$)              &  124$\pm$12&  96 \\
   & $\jpsi \KS$ ($\KS \to \pi^0 \pi^0$)             &   18$\pm$4& 91 \\
   & $\psitwos \KS$                                  &   27$\pm$6&  93 \\ 
\hline \hline

Hadronic &   $D^{*-}\pi^+$  & 622$\pm$27 & 90  \\
(neutral)   &   $D^{*-}\rho^+$ & 419$\pm$25 & 84  \\
            &   $D^{*-}a_1^+$  & 239$\pm$19 & 79  \\ 
            &   $D^{-}\pi^+$   & 630$\pm$26 & 90  \\ 
            &   $D^{-}\rho^+$  & 315$\pm$20 & 84  \\ 
            &   $D^{-}a_1^+$   & 225$\pm$20 & 74  \\ 
            &    total         &2438$\pm$57 & 85  \\ 
\hline \hline

Hadronic &   $\Dzb \pi^+$    &  1755$\pm$47& 88 \\
(charged)   &   $\Dstarb \pi^+$ &   543$\pm$27& 89 \\ 
            &    total          &  2293$\pm$54& 88 \\ 
\hline  \hline

Semileptonic  &   $D^{*-}\ell^+\nu$ &   7517$\pm$104   & 84 \\ 
\hline \hline
  
Control    & $\jpsi K^+ $   &  597$\pm$25&  98 \\
        & $\psitwos K^+ $   &   92$\pm$10&  93 \\
        & $\jpsi K^{*0}$ ($K^{*0} \to K^+ \pi^-$)   &  251$\pm$16 & 95 \\ 
\hline
\end{tabular}
\end{center}
\label{tab:CharmoniumYield}
\end{table*}

\par
The time resolution function 
is described accurately by the sum of
 two Gaussian distributions, which has five independent parameters:
\begin{eqnarray}
{\cal {R}}( \, \deltat ; \, \hat {a} \,  ) = \;\;\;\;\;\;\;\;\;\;\;\;\;\;\;\;\;\;\;
\;\;\;\;\;\;\;\;\;\;\;\;\;\;\;\;\;\;\;\;\;\;\;\;\;\;\;\;\;\;\;\;\nonumber\\
 \sum_{i=1}^{2} { \, \frac{f_i}{\sigma_i\sqrt{2\pi}} \, 
{\rm exp} \left(  - ( \deltat-\delta_i )^2/2{\sigma_i }^2   \right) } .\;\;\;\;
\end{eqnarray}
A fit to the time resolution function in
Monte Carlo simulated events indicates 
that most of the events ($f_1 = 1-f_2 = 70\%$) are in the core 
Gaussian, which has a width $\sigma_1 \approx 0.6 \ps$.
The wide Gaussian has a width  $\sigma_2 \approx 1.8\ps$. Tracks from forward-going charm decays included in the reconstruction of the $B_{tag}$ vertex 
introduce a small bias, 
$\delta_1 \approx -0.2 \ps$, for the core Gaussian.  
\par
A small fraction of events have very large values of \deltaz, 
mostly due to vertex reconstruction
problems.  This is  accounted for in the 
parametrization of the time resolution function by a very wide unbiased Gaussian with fixed width of $8\ps$.   
The fraction of events populating this component of the resolution function,   $f_{w}$, is 
estimated  from Monte Carlo simulation as $\sim 1\%$. 
\par 
In likelihood fits, we use the error $\sigma_{\deltat}$ on \deltat\ that is
calculated from the fits to the two \B\ vertices for each individual event.
However, we introduce two scale factors ${\cal S}_1$ and ${\cal S}_2$ 
for the width of the narrow and the wide Gaussian distributions
($\sigma_1={\cal S}_1 \times \sigma_{\deltat}$ and 
$\sigma_2={\cal S}_2 \times \sigma_{\deltat}$) to account for the fact
that the uncertainty on \deltat\ is underestimated due to effects such as
the inclusion of particles from $D$ decays and possible underestimation 
of the amount of material traversed by the particles.
The scale factor ${\cal S}_1$ and the bias $\delta_1$ of the 
narrow Gaussian are free parameters in the fit.  
The scale factor ${\cal S}_2$ and the fraction of events in the wide
Gaussian, $f_2$, are fixed to the values estimated 
from Monte Carlo  simulation by a fit to the pull distribution (${\cal S}_2=2.1$
and $f_2=0.25$). The bias of the wide Gaussian, $\delta_2$, 
is fixed at $0 \ps$. The remaining set of three parameters:
\begin{eqnarray} 
\hat {a} &=& \{ \, {\cal S}_1, \,  \delta_1, \,  f_{w}  \}
\end{eqnarray}
are determined from the observed vertex distribution in data.
\par
Because the time resolution is dominated by the precision 
of the $B_{tag}$ vertex position, we find  no significant 
differences in the Monte Carlo simulation of 
the resolution function parameters 
for the various fully reconstructed decay modes, validating our 
approach of determining the resolution function parameters  $\hat {a}$
with the relatively high-statistics fully-reconstructed \Bz\ data samples,  
and fixing these parameters in the likelihood fit for the 
determination of \stwob\ with the low-statistics \CP\ sample.  
The differences in the resolution function parameters in the 
different tagging categories are also small.  
\par
Table~\ref{tab:Resolution} presents the values of the parameters
obtained from a fit to the hadronic \Bz\ sample. These values are
used in the final fit for \stwob.

\begin{table}[!htb]
\caption{
Parameters of the resolution function determined from the sample
of events with fully-reconstructed hadronic $B$ candidates.
} 
\vspace{0.3cm}
\begin{center}
\begin{tabular}{|cc|cl|} \hline
   \multicolumn{2}{|c|}{Parameter} & \multicolumn{2}{c|} {Value}    \\ \hline \hline
 $\delta_1$  & (\ps)    & $-0.20\pm0.06$  & from fit     \\
 ${\cal S}_1$&   & $1.33\pm0.14$       & from fit     \\
 $f_{w}$       & (\%)  & $1.6\pm0.6$     & from fit     \\
 $f_1$       & (\%)  & $75$              & fixed        \\
 $\delta_2$  & (\ps)  & $0$             & fixed        \\
 ${\cal S}_2$ &  & $2.1$               & fixed        \\
\hline
\end{tabular}
\end{center}
\label{tab:Resolution}
\end{table}

\section{\boldmath \B\ flavor tagging}
\par
Each event with a \CP\ candidate is assigned a $\Bz$ or $\Bzb$ tag if 
the rest of the event ({\it i.e.,} with the daughter
tracks of the $B_{CP}$ removed) satisfies the criteria for one of several 
tagging categories. 
The figure of merit for each tagging category is the effective tagging efficiency  
$Q_i = \varepsilon_i \, \left( 1 - 2\mistag_i \right)^2$, where $\varepsilon_i$ 
is the fraction of events assigned to category $i$ and 
$\mistag_i$ is the probability of misclassifying the tag as a $\Bz$ or $\Bzb$
for this category. $\mistag_i$ is called the mistag fraction.
The statistical error
on \stwob\ is proportional to $1/\sqrt{Q}$, where $Q = \sum_i Q_i$. 
\par
Three tagging categories
rely on the presence of a fast lepton and/or 
one or more charged kaons in the event.  
Two categories, called neural network categories, are based upon
the output value of a neural network algorithm applied to events 
that have not already been assigned to lepton or kaon  
tagging categories.
\par
In the following, 
the tag refers to the $B_{tag}$ candidate.  
In other words, 
a \Bz\ tag indicates that
the $B_{CP}$ candidate was in a \Bzb\ state 
at $\deltat=0$; a \Bzb\ tag indicates that the $B_{CP}$ candidate was in a 
\Bz\ state.
\subsection{Lepton and kaon tagging categories}
\par
The three lepton and kaon categories are called {\tt Electron}, {\tt Muon} and {\tt Kaon}.  
This tagging technique relies on the correlation 
between the charge of a primary lepton from a semileptonic decay or
the charge of a kaon, and the flavor of the decaying $b$ quark.  
A requirement 
on the center-of-mass momentum of the lepton reduces
contamination from low-momentum  opposite-sign leptons coming from 
charm semileptonic decays.  No similar 
kinematic quantities can be used to discriminate against contamination from 
opposite-sign kaons. 
Therefore,  for kaons the optimization of $Q$ relies principally  on 
the balance between kaon identification efficiency and the purity of the kaon 
sample.  
\par
The first two categories, {\tt Electron} and {\tt Muon},  
require the presence of at least one identified lepton (electron or muon) with a
center-of-mass momentum greater than 1.1\gevc.  
The momentum cut  rejects the 
bulk of wrong-sign leptons from charm semileptonic decays. The value is chosen 
to maximize the effective tagging efficiency $Q$.
The tag is  \Bz\ for a positively-charged lepton, \Bzb\ for a negatively-charged lepton.
\par
If the event is not assigned to  either the {\tt Electron} or the {\tt Muon} 
tagging categories, the event is assigned to the {\tt Kaon} tagging category
if  
the sum of the charges of all identified kaons in the event,  
${\rm \Sigma} Q_K$, 
is different from zero.
The tag is \Bz\ if ${\rm \Sigma} Q_K$ is positive, \Bzb\ otherwise.
\par 
If both lepton and kaon tags are present and provide inconsistent
flavor tags, the event is rejected from the lepton and kaon tagging
categories.

\subsection{Neural network categories}
\par 
The use of a 
second tagging algorithm
is motivated by the 
potential flavor-tagging power carried by non-identified 
leptons and kaons, correlations between leptons and kaons, 
multiple kaons, softer leptons from charm semileptonic decays, 
soft pions from $D^*$ decays and more generally by the momentum 
spectrum of charged particles from \B\ meson decays.  
One way to exploit the information contained in a set of correlated
quantities is to use multivariate methods such as neural networks.
\par
We define five different neural networks, called feature nets, each with 
a specific goal.  Four of the five feature nets are track-based~: the 
{\tt L} and {\tt LS} feature nets are sensitive to the presence
of primary and cascade leptons, respectively, the {\tt K} feature net 
to that of charged kaons and the  {\tt SoftPi} feature net to that 
of soft pions from $D^*$ decays.  In addition, the  {\tt Q} feature net 
exploits the charge of the fastest particles in the event.

\par
The variables used as input to the  
neural network tagger are the highest values of 
the {\tt L}, {\tt LS} and {\tt SoftPi} feature net outputs multiplied by the charge, 
the highest and the second highest value of the {\tt K} feature net 
output multiplied by the charge,
and the output of the {\tt Q} feature net.
\par
The output of the neural network tagger, $x_{NT}$, can be mapped onto the interval $\left[ -1, 1 \right]$.  The tag is \Bz\ if $x_{NT}$ is negative, \Bzb\ otherwise.
Events with $\left| x_{NT} \right| > 0.5$ are classified in the {\tt NT1} tagging category 
and events with  $0.2 < \left| x_{NT} \right| < 0.5$ in the {\tt NT2} tagging category.
Events with  $\left| x_{NT} \right| < 0.2$  have very little tagging power and are excluded from the sample used in the analysis.

\section{Measurement of mistag fractions}
\par
The mistag fractions are measured directly
in events 
in which one \Bz\ candidate, called the $B_{rec}$, is fully reconstructed 
in a flavor eigenstate mode.
The flavor-tagging algorithms described in the previous section 
are applied to the 
rest of the event, which constitutes the potential $B_{tag}$.
\par
Considering the \BzBzb\ system as a whole, one can classify the tagged events 
as {\em mixed} or {\em unmixed} depending on whether the $B_{tag}$ is tagged 
with the same flavor as
the $B_{rec}$ or  with the opposite flavor.   Neglecting the effect of 
possible background contributions,  and assuming the $B_{rec}$ is 
properly tagged,
one can express the measured time-integrated fraction of 
mixed events $\chi$ as a function of the precisely-measured \BzBzb\ mixing probability  $\chi_d$~:
\begin{equation}
\label{eq:TagMix:Integrated}
\chi = \chi_d  + (1 - 2 \chi_d )\, \mistag
\end{equation}
where
$\chi_d = \frac{1}{2} \, x_d^2 / ( 1+ x_d^2 ) $, 
with $ x_d = \deltamd /
\Gamma $.
Thus one can deduce 
an experimental value of the mistag fraction $\mistag$
from the data.
\par
\begin{figure}[t]
\begin{center}
\epsfxsize2.7in
\figurebox{}{}{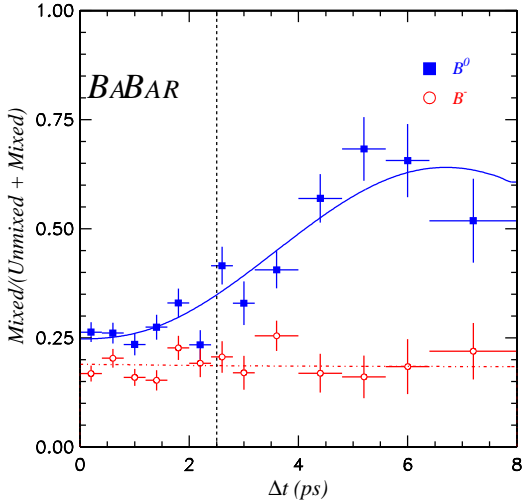}
\caption{Fraction of mixed events $m/(u+m)$ as a function of $|\deltat|$ (ps) for data 
events in the hadronic sample, for neutral $B$ mesons (full squares)
and charged $B$ mesons (open circles). All tagging categories are included. 
This rate is a constant as a function of $\deltat$ for charged $B$ mesons, 
but develops a mixing oscillation for neutral $B$ mesons. 
The rate of mixed events extrapolated to $\deltat=0$ is governed by the mistag 
fraction $\mistag$.  The dot-dashed line at $t_{cut}=2.5 \ps$ indicates the 
bin boundary of the time-integrated single-bin method.}
\label{fig:mistagData}
\end{center}
\end{figure}

A time-dependent analysis of the fraction of mixed events is even more 
sensitive to the mistag fraction.  
The mixing probability is smallest for small values of $\deltat = t_{rec}-t_{tag}$ 
so that the apparent rate of mixed events near \deltat=0 
is governed by the mistag probability (see Fig.~\ref{fig:mistagData}).
A time-dependent analysis can also help discriminate against backgrounds with 
different time dependence.

By analogy with Eq.~\ref{eq:Convol},
we can
express the density functions for unmixed $(+)$ and mixed $(-)$ events as
\begin{eqnarray}
{\cal H}_\pm(\, \deltat; \, \Gamma, \, \deltamd, \, {\cal {D}}, \, \hat {a} \, )  =
\;\;\;\;\;\;\;\;\;\;\;\;\;\;\;\;\;\;\;\;\;\;\;\; \nonumber\\
  \, h_\pm( \deltat;\, \Gamma, \, \deltamd, \, {\cal {D}} \, ) \otimes 
{\cal {R}}( \, \deltat; \, \hat {a} )  ,\;\;\;\;\;
\end{eqnarray}
where
\begin{eqnarray}
h_\pm( \, \deltat; \, \Gamma, \, \deltamd, \, {\cal {D}} \, )  = 
\;\;\;\;\;\;\;\;\;\;\;\;\;\;\;\;\;\;\;\;\;\;\;\;\;\;\;\;\; \nonumber\\
{\frac{1}{4}}\, \Gamma \, {\rm e}^{ - \Gamma \left| \deltat \right| }
\, \left[  \, 1 \, \pm \, {\cal {D}} \times \cos{ \deltamd \, \deltat } \,  \right] .
\;\;\;
\end{eqnarray}
These functions are used to build the log-likelihood function for the mixing analysis:
\begin{eqnarray}
 & \ln { {\cal {L} }_{M} } = \;\;\;\;\;\;\;\;\;\;\;\;\;\;\;\;\;\;\;\;\;\;\;\;\;\;\;\;\;\;\;
\;\;\;\;\;\;\;\;\;\;\;\;\;\;\;\;\;\;\;\;\;\;\;\;\;\;\;\;      \nonumber \\
 & \sum_{ i} \bigg[ \, \sum_{{\rm unmixed}}{  \ln{ {\cal H}_+( \, t;  
\, \Gamma, \, \deltamd, \,   \hat {a},  
\, {\cal {D}}_i \,) } } \;\;\;\;\;\;\;\;\;\;\; \nonumber\\
 & +  \sum_{ {\rm mixed} }{ \ln{ {\cal H}_-( \, t; 
\, \Gamma, \, \deltamd, \,   \hat {a}, 
\, {\cal {D}}_i  \, ) } } \, \bigg] , \;\;\;\; \nonumber\\
\end{eqnarray}
which is maximized to extract the estimates 
of the mistag fractions $\mistag_i = \frac{1}{2}(1-{\cal D}_i)$.

The extraction of the mistag probabilities for each tagging category 
is complicated by the possible presence of mode-dependent 
backgrounds. 
We deal with these
by adding specific terms in the likelihood 
functions describing the different types of backgrounds (zero lifetime, 
non-zero lifetime without mixing, non-zero lifetime with mixing).
described in Ref.~6. 

A simple time-integrated single bin method is used as 
a check of the time-dependent analysis  
for the determination of dilutions from the 
fully reconstructed \Bz\ sample. 
The mistag fractions are deduced from the number of unmixed events, $u$, and 
the number of mixed events,  $m$, 
in a single optimized \deltat\ interval, $\left| \,  \deltat \, \right| < t_{cut}$. 
The bin boundary $t_{cut}$, chosen to minimize the statistical 
uncertainty on the measurement,  is
equal to 2.5 \ps, {\it i.e.,} $1.6$ \Bz\ lifetimes. 
($t_{cut}$  is indicated by a 
dot-dashed line in Fig.~\ref{fig:mistagData}.)  
The resulting mistag fractions based on this method are 
in good agreement with 
the mistag fractions obtained with the maximum-likelihood fit\cite{BabarPub0008}.

\subsection{Tagging efficiencies and mistag fractions}
\begin{table*}[!htb]
\caption{
Mistag fractions measured from 
a maximum-likelihood fit to the time distribution for the fully-reconstructed \Bz\ sample.
The {\tt Electron} and {\tt Muon}  categories are grouped into 
one {\tt Lepton} category.  
The uncertainties on $\varepsilon$ and $Q$ are statistical only.
} 
\vspace{0.3cm}
\begin{center}
\begin{tabular}{|l|c|c|c|} \hline
Tagging Category & $\varepsilon$ (\%) & $\mistag$ (\%) & $Q$ (\%)       \\ \hline \hline
{\tt Lepton}     & $11.2\pm0.5$ & $9.6\pm1.7\pm1.3$   &  $7.3\pm0.7$  \\
{\tt Kaon}       & $36.7\pm0.9$ & $19.7\pm1.3\pm1.1$  &  $13.5\pm1.2$  \\
{\tt NT1}        & $11.7\pm0.5$ & $16.7\pm2.2\pm2.0$  &  $5.2\pm0.7$  \\
{\tt NT2}        & $16.6\pm0.6$ & $33.1\pm2.1\pm2.1$  &  $1.9\pm0.5$  \\  \hline \hline
all              & $76.7\pm0.5$ &                     &  $27.9\pm1.6$ \\ 
\hline
\end{tabular}
\end{center}
\label{tab:TagMix:mistag}
\end{table*}

\par
The mistag fractions and the tagging efficiencies are summarized 
in Table~\ref{tab:TagMix:mistag}. We find a tagging efficiency of 
$(76.7\pm0.5)\%$ (statistical error only).
The lepton categories have the lowest mistag fractions, 
but also have low efficiency.  
The {\tt Kaon} 
category, despite having a larger mistag fraction (19.7\%),  has a higher effective tagging efficiency; one-third of events are assigned to this category.   
Altogether, lepton and kaon categories have an effective tagging 
efficiency $Q \sim 20.8\%$.  
Most of the separation into \Bz\ and  \Bzb\ in the {\tt NT1} and {\tt NT2} tagging categories
derives from the {\tt SoftPi} and {\tt Q} feature nets.  Simulation studies 
indicate that roughly 40\% of the effective tagging efficiency occurs in events that
contain a soft $\pi$ aligned with the $B_{tag}$ thrust axis,  25\% from 
events which have a track with $p^{*} > 1.1 \gevc$, 10\%  from events which 
contain multiple leptons or kaons with opposite charges and are thus not 
previously used in tagging, and the remaining 25\%  from a mixture of the 
various feature nets.  The neural network categories increase the 
effective tagging efficiency by $\sim 7\%$ to an overall 
$Q = (27.9 \pm 1.6) \%$ (statistical error only).

\begin{table*}[!htb]
\caption{
Categories of tagged events in the \CP\ sample.
} 
\vspace{0.3cm}
\begin{center}
\begin{tabular}{|l||c|c|c||c|c|c||c|c|c||c|c|c|} \hline
  & \multicolumn{6}{|c||}{\rule[-1pt]{0mm}{14pt}$\jpsi \KS$} 
  & \multicolumn{3}{|c||}{$\psitwos \KS$} 
  & \multicolumn{3}{|c|}{\CP\ sample } \\ \cline{2-13}
Tagging Category 
  &  \multicolumn{3}{|c||} {\rule[-1pt]{0mm}{14pt}($\KS \to \pi^+\pi^-$)} 
  &  \multicolumn{3}{|c||} {($\KS \to \pi^0\pi^0$)} 
  &  \multicolumn{3}{|c||} {($\KS \to \pi^+\pi^-$)} 
  &  \multicolumn{3}{|c|}  {(tagged)} 
  \\ \cline{2-13}
     &  \Bz\ & \rule[-1pt]{0mm}{14pt}\Bzb\ & all &  \Bz\ & \Bzb\ & all &  \Bz\ & \Bzb\ & all &  \Bz\ &
 \Bzb\ & all \\ \hline \hline 
{\tt Electron}   
  & 1  &  3 & 4  
  & 1  &  0 & 1  
  & 1  &  2 & 3 
  & 3  &  5 & 8      \\
{\tt Muon}       
  & 1  &  3 & 4  
  & 0  &  0 & 0  
  & 2  &  0 & 2 
  & 3  &  3 & 6      \\
{\tt Kaon}       
  & 29 & 18 & 47 
  & 2  & 2  & 4  
  & 5  & 7  &12 
  & 36 & 27 & 63     \\
{\tt NT1}       
  &  9 &  2 & 11 
  & 1  & 0  & 1  
  & 2  & 0  & 2 
  & 12 & 2  & 14     \\
{\tt NT2}       
  & 10 &  9 & 19 
  & 3  & 3  & 6  
  & 3  & 1  & 4 
  & 16 & 13 &  29    \\
\hline \hline
{\tt Total}      
  &  50& 35 & 85 
  & 7  & 5  &12 
  &13  &10  &23  
  & 70 & 50 & {\bf 120 }   \\

\hline
\end{tabular}
\end{center}
\label{tab:TaggedEvents}
\end{table*}

\par
Of the 168 \CP\ candidates, 120 are tagged:  70 as \Bz\ and 50 as \Bzb.
The number of tagged events per category is given in Table~\ref{tab:TaggedEvents}.

\section{Systematic uncertainties and cross checks}
\par
Systematic errors arise from uncertainties in 
input parameters to the maximum likelihood fit, 
incomplete knowledge of the time resolution function, 
uncertainties in the mistag fractions, 
and possible limitations in the analysis procedure.  
We fix the \Bz\ lifetime to the nominal PDG\cite{PDG2000} central value $\tau_{\Bz} = 1.548\ \ps$ 
and the value of ${\rm \Delta} m_d $ to the nominal PDG value $ {{\rm \Delta} m_d } = 0.472 \, \hbar \ps^{-1}$.
The errors on \stwob\ due to uncertainties in  
$\tau_{\Bz}$ and \deltamd\ are 0.002 
and 0.015, respectively.
The remaining systematic uncertainties are discussed in the following sections.

\subsection{Uncertainties in the resolution function}
\par
The time resolution is measured with the high-statistics sample of 
fully-reconstructed \Bz\ events.  The time  resolution for the 
\CP\ sample should be very similar, especially to that measured 
for the 
hadronic sample. 
We verify that the resolution function extracted in
the hadronic sample is consistent with the one extracted in the semileptonic sample.
We assign as a systematic error the  variation in \stwob\ 
obtained by changing the resolution
parameters by one statistical standard deviation. 
The corresponding error on \stwob\
is 0.019.
\par
We use a full Monte Carlo simulation to verify that 
the Bremsstrahlung recovery procedure in the $\jpsi \to \epem$ mode does not
introduce any systematic bias in the \deltat\  measurement, 
nor does it affect the vertex resolution and pull distributions.
\par
In order to check the impact of imperfect knowledge
of the bias in \deltat\ on the measurement, we allow 
the bias of the second Gaussian to increase to 0.5 \ps.
The resulting change in \stwob\ of  0.047 is assigned as 
a systematic error. 
The sensitivity to the bias is due to the different
number of events tagged as  \Bz\ and  \Bzb.
 
\subsection{Uncertainties in flavor tagging}
\par
The mistag fractions are measured with uncertainties that are  either 
correlated 
or uncorrelated between tagging categories.  
We study the effect of uncorrelated
errors (including statistical errors)  on the asymmetry
by varying the mistag fractions individually for each category, using
the full covariance matrix.
For correlated errors, we vary the mistag fractions 
for all categories simultaneously.
\par
The main common source of systematic uncertainties in the 
measurement of mistag fractions is the presence of backgrounds, 
which are more significant in the semileptonic sample 
than in the hadronic sample.
The largest background is due to random combinations of particles  
and can be studied with mass sidebands.  
Additional backgrounds arise in the 
semileptonic sample from misidentified leptons, from leptons 
incorrectly associated with a true $D^{*}$ from $B$ decays, 
and from charm events 
containing a   $D^{*}$ and a lepton.  
The details of the procedure for accounting for the backgrounds
and the uncertainties on the background levels,
and the estimates of resulting
systematic errors on the mistag fractions are given in Ref.~6.
We estimate the systematic error on \stwob\ due to the uncertainties in the measurement
of the mistag fractions to be 0.053, for our \CP\ sample.
\par
In the likelihood function, we use the same mistag fractions 
for the \Bz\ and \Bzb\ samples.  
However, differences are expected due to effects such as the  
different cross sections for $K^+$ and $K^-$ hadronic interactions.
For equal numbers of tagged \Bz\ and \Bzb\ events, the impact on \stwob\
of a difference in 
mistag fraction,   ${\rm \delta} \mistag \! = \! \mistag_{\Bz} \! - \! \mistag_{\Bzb}$, is insignificant.
From studies of charged and neutral \B\ samples, we find that
the mistag differences  are $\leq 0.02$  for the {\tt NT1} category,  
$\leq 0.04$ for the {\tt Kaon} category, and negligible for the 
lepton categories. 
However, for the {\tt NT2} category, there is a significant 
difference between the \Bz\ and \Bzb\ mistag fractions, 
$\delta \mistag = 0.16$, which 
is not predicted by our simulation. 
Although this would lead to 
a negligible systematic shift in \stwob, we cover the possibility 
of different mistag fractions in the \CP\ sample and 
the fully-reconstructed sample used to measure the mistag fractions
by assigning as a systematic uncertainty the shift in
\stwob\ resulting from using the measured mistag fraction 
for the {\tt NT2}
category from the sample of $\jpsi K^{*0}$  events only.  
The resulting conservative systematic 
uncertainty on \stwob\ is 0.050.
\par  
For a small sample of events, 
there can be a significant difference in the number of 
\Bz\ and \Bzb\ events, $\Delta N = N_{\Bz}-N_{\Bzb}$.  
For a single tagging category, the fractional change in \stwob\ 
from such a difference is 
$\, \Delta \stwob / \stwob \approx  \delta \mistag  \Delta N /  N$.
In the \CP\ sample, 
$\Delta  N /  N$ is significant  only in the {\tt Kaon} and {\tt NT1} categories 
(see Table~\ref{tab:TaggedEvents}).  
Taking into account their relative weight 
in the overall result, we assign a fractional systematic error 
of $0.005$
on \stwob. 
\par 
The systematic uncertainties on the mistag fractions due to 
the uncertainties on $\tau_{\Bz}$ and \deltamd\ are negligible.

\subsection{Uncertainties due to backgrounds}
\par
The fraction of background events in the  \CP\ sample  
($\jpsi \KS$ and $\psitwos \KS$)
is estimated to be $(5\pm3)\%$.  
The portion of this
background that occurs at small values of \deltat\ ({\it e.g.,}
contributions from $u$, $d$ and $s$ continuum events)  does not contribute 
substantially to the determination of the asymmetry.    We estimate that this
reduces the effective background to $3\%$.
We correct for the
background by increasing the apparent asymmetry by a factor of $1.03$. 
In addition, 
we assign a fractional systematic uncertainty of $3\%$ on the asymmetry,
to cover both the uncertainty in the size of the background and
the possibility that the background might have some \CP-violating component.

\section{Extracting \boldmath \stwob}
\subsubsection{Blind analysis}
\par
We have adopted a blind analysis for the extraction of \stwob\ 
in order to eliminate possible experimenter's bias.  
We use a technique 
that hides not only the result of the unbinned 
maximum likelihood fit, but also the visual \CP\ asymmetry
in the \deltat\ distribution.  
The error on the asymmetry is not hidden.

\begin{figure}[!h]
\begin{center}
\epsfxsize2.7in
\figurebox{}{}{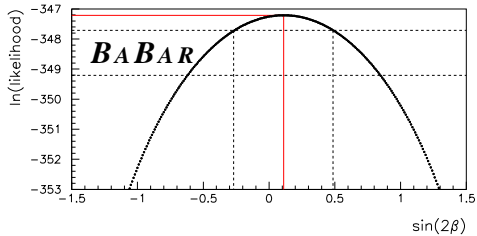}
\caption{Variation of the log likelihood as a function of \stwob. 
The two horizontal dashed lines indicate changes in the log likelihood 
corresponding to one and two statistical standard deviations.}
\label{fig:likelihood}
\end{center}
\end{figure}

The amplitude of the asymmetry ${\cal A}_{CP}(\deltat)$
from the fit 
is hidden from the experimenter
by arbitrarily flipping its sign and adding an arbitrary offset.
The sign flip hides whether a change in the analysis 
increases or decreases the resulting asymmetry.
However, the magnitude of the change is not hidden. 

The visual \CP\ asymmetry in the  \deltat\ distribution is  
hidden by multiplying \deltat\ by the sign of the tag 
and  adding an arbitrary offset.

With these techniques, systematic studies can be performed
while keeping the numerical value of \stwob\  hidden.   In particular, 
we can check that the hidden \deltat\ distributions are 
consistent for \Bz\ and \Bzb\ tagged events. The same is true for all the 
other checks concerning tagging, vertex resolution and the 
correlations between them.   
For instance, fit results in the different tagging 
categories can be compared to each other, since each fit is 
hidden in the same way. The analysis procedure for extracting 
\stwob\ was frozen, and the
data sample fixed, prior to unblinding. 

\begin{figure}[t]
\begin{center}
\epsfxsize2.7in
\figurebox{}{}{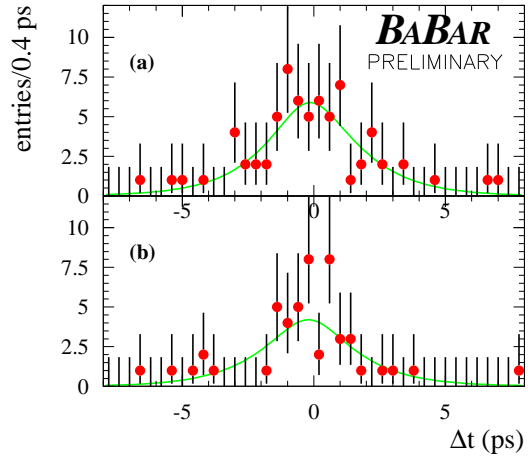}
\caption{Distribution of \deltat\ for (a) the \Bz\ tagged events and (b) the  \Bzb\ tagged events 
in the \CP\ sample.  The error bars plotted for each data point assume 
Poisson statistics.  
The curves correspond to the result of the unbinned maximum-likelihood fit
and are each normalized to the observed number of tagged \Bz\ or \Bzb\ events.}
\label{fig:deltatfit}
\end{center}
\end{figure}

\begin{figure}[t]
\begin{center}
\epsfxsize2.7in
\figurebox{}{}{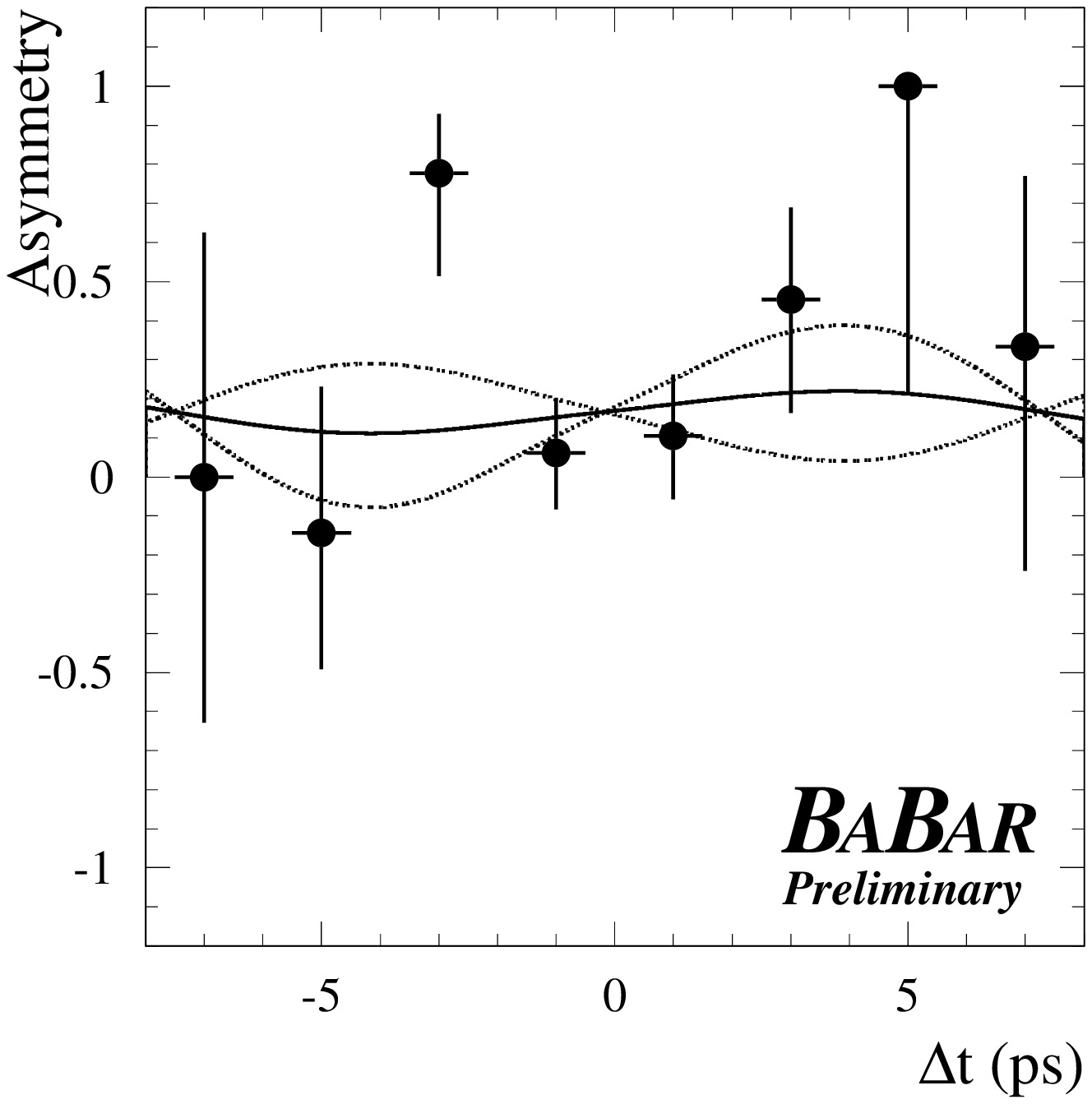}
\caption{The raw \Bz-\Bzb\ 
asymmetry $(N_{\Bz}-N_{\Bzb})/(N_{\Bz}+N_{\Bzb})$, with binomial errors,  
is shown as a function of \deltat.  
The time-dependent asymmetry is represented by a solid curve 
for our central value of \stwob, 
and by two dotted curves for the values at plus and minus 
one statistical standard deviation from the central value.
The curves are not centered at $(0,0)$ in part because 
the probability density functions are normalized 
separately for \Bz\ and \Bzb\ events, and
our \CP\ sample contains an unequal number of 
\Bz\ and \Bzb\ tagged events (70 \Bz\ versus 50 \Bzb).
The $\chi^2$ between the binned asymmetry and the result of the maximum-likelihood
fit is 9.2 for 7 degrees of freedom.}
\label{fig:asymmetry}
\end{center}
\end{figure}

\subsection{Cross checks of the fitting procedure}

We submitted our maximum-likelihood fitting procedure 
to an extensive series of simulation tests. The tests were carried
out with two different implementations of the fitting algorithm 
to check for software errors.
The validation studies were done on two types of simulated event samples.

\begin{itemize}
\item
``Toy'' Monte Carlo simulation tests.  
In these samples, the detector response is not simulated.
Monte Carlo techniques are used with parametrized resolution functions
and tagging probabilities.
We validated the fitting procedure on large samples of simulated 
\CP\ events, for various numbers of tagging categories,
values of mistag fractions and values of \stwob. 
We also simulated a large number of 100-event experiments, 
with the purpose of investigating statistical issues 
with small samples, including values of \stwob\ near unphysical regions.   
We checked that the fitter performs well in the presence
of backgrounds for the extraction of the mistag fractions.  
We exercised the combined \CP\ and mixing fits, and found that
although combined fits perform well, 
they do not significantly improve the statistical sensitivity of the result.

\item 
Full Monte Carlo simulation tests.  We studied samples of
$\jpsi \KS$,  $\jpsi K^+$, $D^*\pi$ and $D^*\ell\nu$ events 
produced with the \babar\ 
GEANT3 detector simulation and reconstructed with the \babar\ reconstruction
program. 
$\jpsi \KS$ events were generated with various values of \stwob.  
We  extracted the ``apparent 
\CP-asymmetry'' for the charged \B's and found it to be consistent
with zero. 
We studied the difference
in tagging efficiencies and in mistag fractions between the 
charged and neutral $B$ samples.  We also tested
the procedure for extracting the mistag fractions from 
hadronic and semileptonic samples of 
fully simulated events ($D^*\pi$ and $D^*\ell\nu$).
\end{itemize}

\section{Results}
\begin{table}[!t]
\caption{
Result of fitting for \CP\ asymmetries in the entire \CP\ sample and in 
various subsamples.
} 
\vspace{0.3cm}
\begin{center}
\begin{tabular}{|l|c|} \hline
 sample                                    &  \stwob  \\ \hline \hline
 \ \ \CP\ sample                               &  {\bf 0.12}$\pm${\bf 0.37}  \\  
\hline
 \ \ $\jpsi \KS$ ($\KS \to \pi^+ \pi^-$) &  $-0.10 \pm 0.42$   \\  
 \ \ other \CP\ events                           &  $0.87 \pm 0.81$   \\  
\hline
 \ \ {\tt Lepton}                         &  $1.6 \pm 1.0  $  \\
 \ \ {\tt Kaon}                           &  $0.14\pm 0.47   $   \\
 \ \ {\tt NT1}                            &  $-0.59\pm0.87  $   \\
 \ \ {\tt NT2}                            &  $-0.96\pm 1.30  $   \\
\hline 
\end{tabular}
\end{center}
\label{tab:result}
\end{table}

\begin{table*}[!t]
\caption{
Summary of systematic uncertainties.  
We compute the 
fractional systematic errors
using the actual value of our asymmetry 
increased by one statistical standard deviation, 
that is $0.12+0.37 = 0.49$.  
The different contributions to the systematic 
error are added in quadrature.
} 
\vspace{0.3cm}
\begin{center}
\begin{tabular}{|l|c|} \hline
 Source of uncertainty    &  Uncertainty on \sb \\ \hline \hline
 
 Uncertainty on $\tau_\Bz$                         &   0.002     \\
 Uncertainty on \deltamd                           &   0.015     \\
 Uncertainty on \deltaz\ resolution for \CP\ sample  &   0.019     \\ 
 Uncertainty on time-resolution bias for \CP\ sample               &   0.047     \\ 
 Uncertainty on measurement of mistag fractions        &  0.053    \\   
 Different mistag fractions for \CP\ and non-\CP\ samples       &   0.050     \\
 Different mistag fractions for \Bz\ and \Bzb\     &   0.005     \\
 Background in \CP\ sample                         &   0.015     \\
\hline \hline
 Total systematic error                            & {\bf 0.091 }   \\ 
\hline 
\end{tabular}
\end{center}
\label{tab:systematics}
\end{table*}

\par
The maximum-likelihood fit for \stwob, using the full tagged sample of $\Bz/\Bzb \to \jpsi \KS$ and $\Bz/\Bzb \to \psitwos \KS$ 
events, gives:
\begin{eqnarray}
\stwob=0.12\pm 0.37 {\rm (stat)} \pm 0.09 {\rm (syst)}\, . \nonumber
\end{eqnarray}
For this result, the \Bz\ lifetime and \deltamd\ are
fixed to the current best values\cite{PDG2000},
and \deltat\ resolution parameters and the mistag rates are fixed to the values obtained from data as summarized
in Tables~\ref{tab:Resolution} and \ref{tab:TagMix:mistag}.
The log likelihood is shown as a function of \stwob\ 
in Fig.~\ref{fig:likelihood}, 
the \deltat\ distributions
for \Bz\ and \Bzb\ tags in Fig.~\ref{fig:deltatfit},
and the raw asymmetry as 
a function of \deltat\ in Fig.~\ref{fig:asymmetry}. 
The results of the fit for each
type of \CP\ sample and for each tagging category are given in Table~\ref{tab:result}.  
The contributions to the systematic uncertainty are summarized
in Table~\ref{tab:systematics}.

\par 
We estimate the probability of obtaining
the observed value of the statistical uncertainty, 0.37, 
on our measurement of \stwob\
by generating a large number of toy Monte Carlo experiments
with the same number of tagged \CP\ events, and distributed in the same 
tagging categories,  as in the \CP\ sample in the data.  
We find that the errors are distributed
around $0.32$ with a standard deviation  of $0.03$,
and that the probability of obtaining 
a value of the statistical error larger than the one we observe is 5\%.
Based on  a large number of full Monte Carlo simulated experiments with the 
same number of events as our data sample, we 
estimate that the probability of finding a lower value of the likelihood than
our observed value is 20\%.

\section{Validating analyses}
To validate the analysis we use the charmonium 
control sample,  composed of $B^+ \to \jpsi K^+$ events
and events with self-tagged  $\jpsi K^{*0}$  ($K^{*0} \to K^+ \pi^-$)
neutral \B's.  We also use the event samples with fully-reconstructed
candidates in charged or neutral hadronic modes.
These samples should exhibit no time-dependent asymmetry. 
In order to 
investigate  this experimentally, we define an ``apparent 
\CP\ asymmetry'',  analogous to \stwob\  in 
Eq.~\ref{eq:asymmetry}, which we extract from the data using an identical
maximum-likelihood procedure. 

The events in the control samples are flavor eigenstates and not
\CP\ eigenstates.
They are used  for testing 
the fitting procedure with the same tagging algorithm as for the \CP\
sample and, in the case of the $B^+$ modes, with self-tagging based on 
their charge.
We also perform the fits for \Bz\ and \Bzb\ (or $B^+$ and $B^-$) events
separately to study possible flavor-dependent systematic effects.    
For the charged \B\ modes, we use mistag fractions measured
from the sample of hadronic charged \B\ decays.

In all fits, including the fits to charged samples, 
we fix the lifetime  $\tau_{\Bz}$ and the oscillation frequency  \deltamd\ 
to the PDG values\cite{PDG2000}.

\begin{table}[!t]
\caption{
Results of fitting for apparent \CP\ asymmetries in various 
charged or neutral flavor-eigenstate \B\ samples. 
} 
\begin{center}
\begin{tabular}{|l|c|} \hline
 Sample                                    & Apparent  \\
 & \CP-asymmetry \\ 
\hline \hline
 Hadronic $B^\pm$ decays                  &   $0.03 \pm 0.07$   \\  
\hline
 Hadronic $B^0$ decays            &   $-0.01 \pm 0.08$  \\  
\hline
 $\jpsi K^+$                               &   $0.13\pm 0.14$    \\ 
\hline 
 $\jpsi K^{*0}$,   &  $ 0.49 \pm 0.26$   \\ 
 \hbox{      } $K^{*0} \to K^+ \pi^-$ & \\
\hline 
\end{tabular}
\end{center}
\label{tab:validation}
\end{table}
The results of a series of validation checks on the control samples
are summarized in Table~\ref{tab:validation}. 

The two high-statistics samples and the $\jpsi K^+$ sample 
give an apparent \CP\ asymmetry consistent with zero.  
The 1.9 $\sigma$ 
asymmetry in the $\jpsi K^{*0}$ 
is interpreted
as a statistical fluctuation.

Other \babar\ time-dependent analyses presented at this Conference 
demonstrate the validity of the novel technique developed for 
use at an asymmetric \BF.  The measurement of the \Bz-\Bzb\ oscillation 
frequency described in Ref.~6 
uses the same 
time resolution function and tagging algorithm as the \CP\ analysis.
Fitting for  \deltamd\ in the maximum-likelihood fit for 
the fully-reconstructed hadronic and semileptonic neutral \B\ decays, 
we measure
\begin{eqnarray}
 & \deltamd = \hskip2in \hfil \nonumber\\
 & 0.512 \pm 0.017 {\rm (stat) } \pm 0.022  {\rm (syst) } \, \hbar \ps^{-1} \ , \nonumber
\end{eqnarray}
which is consistent with the world average\cite{PDG2000} 
$\deltamd = 0.472 \pm 0.017 \, \hbar \ps^{-1}$.
The \Bz\ lifetime measurement described in Ref.~7 
uses the same inclusive 
vertex reconstruction technique as the \CP\ analysis.
We measure
\begin{eqnarray}
  \tau_{\Bz} = 1.506 \pm 0.052 {\rm (stat) } \pm 0.029  {\rm (syst) } \ps \ , \nonumber
\end{eqnarray}
also consistent with the world average\cite{PDG2000}  
$\tau_{\Bz} = 1.548 \pm 0.032 \ps $.

\section{Conclusions and prospects} 

\begin{figure}[!htb]
\begin{center}
\epsfxsize2.7in
\figurebox{}{}{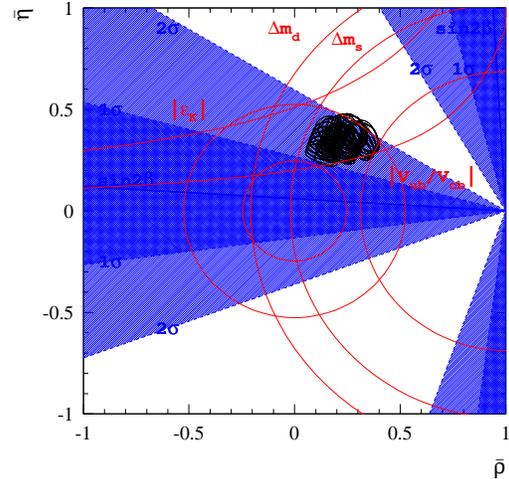}
\caption{
Present constraints on the position of the apex of the Unitarity Triangle in the 
($\bar{\rho},\bar{\eta})$ plane. The fitting procedure is described in Ref.~2. 
Our result $\stwob = 0.12\pm 0.37{\rm (stat)}$ is represented by cross-hatched regions 
corresponding to 
one and two statistical standard deviations.}
\label{fig:UnitarityTriangle}
\end{center}
\end{figure}

We have presented \babar's first measurement of the \CP-violating 
asymmetry parameter \stwob\ in the $B$ meson system:
\begin{eqnarray}
\stwob=0.12\pm 0.37 {\rm (stat)} \pm 0.09 {\rm (syst)}\, . \nonumber
\end{eqnarray} 
Our measurement is consistent with the world average\footnote{Based
on the OPAL result\cite{OPAL} $\stwob = 3.2 ^{+1.8} _{-2.0} \pm 0.5$ and the  
CDF result\cite{CDF}  $\stwob = 0.79^{+0.41}_{-0.44}$. See also ALEPH's preliminary 
result\cite{ALEPH} $\stwob = 0.93^{+0.64 \, +0.36 }_{-0.88 \, -0.24}$.} 
$\stwob = 0.9\pm0.4$, 
and is currently limited by the size of the \CP\ sample.
We expect to more than double the present data sample in the near future.

Figure~\ref{fig:UnitarityTriangle} shows the Unitarity Triangle 
in the ($\bar{\rho},\bar{\eta})$ plane, with \babar's measured central value of 
\stwob\ shown as two straight lines;
there is a two-fold ambiguity in deriving a value of $\beta$ from a measurement of \stwob\ . 
Both choices are shown with cross-hatched regions corresponding
to one and two times the one-standard-deviation experimental uncertainty.
The ellipses correspond to the regions allowed by all other
measurements that constrain the Unitarity Triangle.
Rather than make the common, albeit unfounded, assumption that our lack of knowledge of
theoretical quantities, or differences between theoretical models, can be parametrized
(typically as a Gaussian or  flat distribution), 
we have chosen to display the ellipses corresponding to measurement errors at a variety
of representative choices\footnote{We use the
following set of measurements: 
$\left| V_{cb} \right| = 0.0402\pm0.017$, $ \left| V_{ub}/V_{cb} \right| = 
\left< \left| V_{ub}/V_{cb} \right| \right> \pm 0.0079$, $\deltamd = 0.472\pm0.017 \, 
\hbar \ps^{-1}$ and $\left| \epsilon_K  \right| = \left( 2.271 \pm 0.017 \right) \times 10^{-3}$, 
and for ${\rm \Delta} m_s$ the set of amplitudes corresponding to a $95\%$CL limit of $14.6 \, 
\hbar \ps^{-1}$.   
We scan the model-dependent parameters $\left< \left| V_{ub}/V_{cb} \right| \right>$, $B_K$,
$f_{B_d} \sqrt{ B_{B_d} } $ and $\xi_s$, in the range 
$\left[ \, 0.070, \, 0.100 \, \right]$,%
$\left[ \, 0.720, \, 0.980 \, \right]$,  
$\left[ \, 185, \, 255 \, \right] \mev$ and 
$\left[ \, 1.07, \, 1.21 \, \right]$, respectively. } of theoretical parameters.
The fitting procedure is described in Ref.~2.  

While the current experimental uncertainty on \stwob\ is large, the next few years will bring
substantial improvements in precision, as well as measurements 
for other final states in which \CP-violating asymmetries are proportional to \stwob, and
measurements for modes in which the asymmetry is proportional to \stwoa.  

\section{Acknowledgments}

We wish to thank our \pep2\ colleagues for their superb accomplishment in achieving excellent peak
luminosity and remarkable efficiency, enabling us to accumulate a \FourS\ large data sample of excellent
quality in a remarkably short time. \babar\ has received support from the Natural Sciences and Engineering
Research Council (Canada), The Institute of High Energy Physics (China), Commissariat \`a l'Energie Atomique
and Institut National de Physique Nucl{\'e}aire et de Physique des Particules (France), Bundesministerium
f{\"u}r Bildung and Forshung (Germany), Instituto Nazionale di Fisica Nucleare (Italy), The Research
Council of Norway, The Ministry of Science and Technology of the Russian Federation, The Particle Physics
and Astronomy Research Council (United Kingdom) and the US Department of Energy and National Science
Foundation.

\vfill\eject

\end{document}